\pdfoutput=1
\documentclass[twocolumn]{aastex62}
\usepackage{lineno}
\usepackage{amsmath}
\usepackage{empheq}
\usepackage{appendix}
\usepackage{svg}
\usepackage{graphicx}
\usepackage{natbib}
\bibpunct{}{}{,}{a}{}{;}
\shorttitle{}
\shortauthors{Ley et al.}

\begin{document}
\title{A Heating Mechanism via Magnetic Pumping in the Intracluster Medium}
\shorttitle{ICM Heating via Magnetic Pumping}
\correspondingauthor{Francisco Ley}
\email{fley@astro.wisc.edu}
\author[0000-0002-8820-8177]{Francisco Ley}
\affiliation{Department of Astronomy, University of Wisconsin-Madison, Madison, Wisconsin 53706, USA}
\author[0000-0003-4821-713X]{Ellen G. Zweibel}
\affiliation{Department of Astronomy, University of Wisconsin-Madison, Madison, Wisconsin 53706, USA}
\affiliation{Department of Physics, University of Wisconsin-Madison, 1150 University Avenue, Madison,WI, USA 53706}
\author[0000-0003-2928-6412]{Mario Riquelme}
\affiliation{Departamento de F\'{i}sica, Facultad de Ciencias F\'{i}sicas y Matem\'{a}ticas, Universidad de Chile, Chile}
\author[0000-0003-2928-6412]{Lorenzo Sironi}
\affiliation{Department of Astronomy, Columbia University, New York, NY 10027, USA}
\author{Drake Miller}
\affiliation{Department of Astronomy, University of Wisconsin-Madison, Madison, Wisconsin 53706, USA}
\author[0000-0003-3483-4890]{Aaron Tran}
\affiliation{Department of Astronomy, Columbia University, New York, NY 10027, USA}


\begin{abstract}
Turbulence driven by AGN activity, cluster mergers and galaxy motion constitutes an attractive energy source for heating the intracluster medium (ICM). How this energy dissipates into the ICM plasma remains unclear, given its low collisionality and high magnetization (precluding viscous heating by Coulomb processes). \citet{Kunz2011} proposed a viable heating mechanism based on the anisotropy of the plasma pressure (gyroviscous heating) under ICM conditions. The present paper builds upon that work and shows that particles can be gyroviscously heated by large-scale turbulent fluctuations via magnetic pumping. We study how the anisotropy evolves under a range of forcing frequencies, what waves and instabilities are generated and demonstrate that the particle distribution function acquires a high energy tail. For this, we perform particle-in-cell simulations where we periodically vary the mean magnetic field $\textbf{B}(t)$. When $\textbf{B}(t)$ grows (dwindles), a pressure anisotropy $P_{\perp}>P_{\parallel}$ ($P_{\perp}< P_{\parallel}$) builds up ($P_{\perp}$ and $P_{\parallel}$ are, respectively, the pressures perpendicular and parallel to $\textbf{B}(t)$). These pressure anisotropies excite mirror ($P_{\perp}>P_{\parallel}$) and oblique firehose ($P_{\parallel}>P_{\perp}$) instabilities, which trap and scatter the particles, limiting the anisotropy and providing a channel to heat the plasma. The efficiency of this mechanism depends on the frequency of the large-scale turbulent fluctuations and the efficiency of the scattering the instabilities provide in their nonlinear stage. We provide a simplified analytical heating model that captures the phenomenology involved. Our results show that this process can be relevant in dissipating and distributing turbulent energy at kinetic scales in the ICM. 
\end{abstract} 

\submitjournal{ApJ}
\section{Introduction}
\label{sec:intro}

The intracluster medium (ICM) of galaxy clusters strongly affects the structure and evolution of the embedded galaxies and is of cosmological significance. It is therefore critical to understand the dynamics and energy balance of the ICM.

The observed X-ray emission from the ICM implies radiative cooling timescales shorter than their typical ages (e.g. \cite{Voigt2004}). In the absence of a heat source that counteracts this cooling, significant mass inflow takes place (\cite{FabianNulsen1977}), providing a continuous supply of cool gas that could fuel star formation. Further X-ray observational studies have inferred much lower inflow mass rates (\cite{Peterson2003}) and star formation rates (\cite{Donahue2015}) than predicted. This conundrum is well known as the cooling flow problem. 

Among the various heating mechanisms that have been proposed to counteract the radiative cooling of the ICM (see \cite{Zweibel2018} for a review), active galactic nuclei (AGN) feedback has been considered promising in terms of, e.g., the amount of energy injected into the ICM and its self-regulating nature (\cite{Churazov2001,Reynolds2002,Omma2004,Yang2016}). However, it is not well understood how the energy in AGN outflows is ultimately deposited and thermalized in the ICM. 

Turbulent heating has been suggested as a channel of energy dissipation able to offset radiative cooling in the ICM (\cite{Zhuravleva2014}), but the study of turbulent dissipation in the ICM presents several challenges both theoretically and observationally. The well-known presence of $\mu G$ magnetic fields (\cite{Bonafede2010}), its weakly collisional nature (i.e. the typical Coulomb collision time and mean free path are much longer than the ion Larmor period and Larmor radius, and in fact approach dynamical time and length scales), 
make the transport properties of the plasma, such as its thermal conduction and viscosity, anisotropic with respect to the direction of the local magnetic field and strongly dependent on the microphysics happening at kinetic scales. Additionally, the large ratio of thermal to magnetic pressure ($\beta \equiv 8\pi P/B^2 \sim 10-100$ where $P$ is the isotropic thermal pressure and $B$ the magnetic field strength) put ICM plasmas in a regime that is not well studied experimentally or observed in space plasmas.
 
 Microphysical processes driven by plasma pressure anisotropy appear to be particularly important. The compression, rarefaction, and shearing of collisionless, magnetized plasma generally produces
 pressure anisotropy with respect to the local magnetic field ($P_{\perp,j}\neq P_{\parallel,j}$, where $P_{\perp,j}$ and $P_{\parallel,j}$ are, respectively, the pressures of the species $j$ perpendicular and parallel to the magnetic field), as was first pointed out in the context of galaxy clusters by \cite{Schekochihin2005} and \cite{Schekochihin2006}. These authors showed that even a small anisotropy can make the plasma easily unstable to microinstabilities such as mirror ($P_{\perp,j}>P_{\parallel,j}$) and oblique firehose ($P_{\parallel,j}>P_{\perp,j}$), the thresholds for which scale as $\sim \beta^{-1}$. 
 
 A turbulent heating mechanism based on the anisotropy of the ICM plasma was proposed by \cite{Kunz2011}.
 According to this  mechanism, which is known as gyroviscous heating (\cite{Kulsrud1983,Hollweg1985}), the heating rate is determined by the level of pressure anisotropy $\Delta P$ in the system. These authors estimated $\Delta P$ by assuming it is  pinned to marginal stability with respect to the mirror and firehose microinstabilities. A related mechanism, termed magnetic pumping, was explored by \cite{Lichko2017} with application to the solar wind. In that work, the authors considered heating and nonthermal particle energization under the assumption of a periodically compressing flux tube with a prescribed scattering frequency.

Gyroviscous heating and magnetic pumping have the important property that they directly tap the energy of large scale turbulence in weakly collisional plasmas. This is in notable contrast to the usual assumption made, e.g. in hydrodynamic turbulence, that energy cascades from large scales to small in a loss free manner, with dissipation becoming significant only at small scales. 

In this work, we combine the approaches of \cite{Kunz2011} and \cite{Lichko2017} and present a heating mechanism based on gyroviscous heating acting via magnetic pumping in a
periodically shearing, high-$\beta$ plasma suitable for studying ICM conditions. The isotropizing effect necessary for the operation of magnetic pumping will be given by the scattering that mirror and firehose instabilities provide as they interact with the particles, and this will be self-consistently captured using 2.5-dimensional (i.e. electromagnetic fields and particles' momenta have 3 components but only vary in 2 spatial dimensions), fully kinetic particle-in-cell (PIC) simulations. This mechanism allows the plasma to effectively retain energy coming from large-scale turbulent fluctuations after a complete pump cycle. One of the outcomes of our work is that, as predicted by \cite{Lichko2017}, shorter cycles lead to greater energy retention, even when the scattering process is modeled self consistently.

This paper is organized as follows. In \S \ref{sec:TheoreticalModel} we present the analytical basis 
for plasma heating via magnetic pumping by gyroviscosity. In \S \ref{sec:Simulation Setup} we present our simulation method and setup. In \S \ref{sec:Results} we quantify the amount of gyroviscous heating in our magnetic pumping configuration, show that our results are fairly independent of the numerical ion-to-electron mass ratio, and discuss their dependence on the rate at which $B$ is driven. In \S \ref{sec:AnisotropyEvolution} we describe how magnetic pumping operates in terms of the evolution of the pressure anisotropy along with the alternating excitation of mirror and firehose instabilities and discuss their dependence on different physical parameters. In \S \ref{sec:ScatteringModel} we provide a simplified heating model able to capture the essence of the magnetic pumping mechanism in presence of scattering by
self-consistently generated kinetic instabilities. In \S \ref{sec:Conclusions} we summarize our results and present our conclusions.


\section{Theoretical Basis}
\label{sec:TheoreticalModel}

In this section, we develop the analytical basis for capturing the main features of the gyroviscous heating mechanism, particularly how the presence of a scattering mechanism becomes necessary to effectively heat the plasma. 

Consider a uniform plasma subject to an external, incompressible forcing such as the shear velocity field considered in our numerical simulations (cf. section \ref{sec:Simulation Setup}). Consider first the case where there is no scattering mechanism. We can describe the evolution of the particle distribution function $f=f(p_{\perp},p_{\parallel},t)$ by the drift kinetic equation (e.g \cite{Kulsrud1983,Kulsrud2005})
\begin{align}
    \frac{\partial f}{\partial t} + \frac{dp_{\perp}}{dt}\frac{\partial f}{\partial p_{\perp}} + \frac{dp_{\parallel}}{dt}\frac{\partial f}{\partial p_{\parallel}} = 0,
    \label{eq:DriftKineticEquation}
\end{align}
where $p_{\perp}$ and $p_{\parallel}$ are the momenta perpendicular and parallel to the mean magnetic field, respectively. We will assume that the external motion drives the background sufficiently slowly that the particle magnetic moment $\mathcal{M} \equiv p_{\perp}^2/B$ and longitudinal action $\mathcal{J} \equiv \oint p_{\parallel}d\ell \propto p_{\parallel} L$ are adiabatic invariants. For a pure shear motion, assuming conservation of particle number $N=LAn$ and magnetic flux $\Phi = \int \textbf{B}\cdot d\textbf{S} \propto BA$, we can write

\begin{align}
    \frac{d p_{\perp}}{dt} = \frac{p_{\perp}}{2}\frac{\dot{B}}{B}, \ 
    \frac{d p_{\parallel}}{dt} = -p_{\parallel}\frac{\dot{B}}{B},
    \label{eq:MomentumEvolution}
\end{align}
which are valid for both nonrelativistic and relativistic particles.
Substituting eqn. \ref{eq:MomentumEvolution} into eqn. \ref{eq:DriftKineticEquation} gives

\begin{align}
    \frac{\partial f}{\partial t} + \frac{\dot{B}}{B}\left( \frac{p_{\perp}}{2}\frac{\partial f}{\partial p_{\perp}} - p_{\parallel}\frac{\partial f}{\partial p_{\parallel}}\right) = 0.
\end{align}

In order to illustrate the presence of the gyroviscous heating rate and to obtain the evolution of the pressure anisotropy in the most transparent way, it is convenient to do a coordinate transformation from $(p_{\perp},p_{\parallel})$ to $(p,\mu)$, where $p\equiv (p_{\perp}^2 + p_{\parallel}^2)^{1/2}$ and $\mu \equiv p_{\parallel}/p$. The drift kinetic equation in the new coordinate system reads

\begin{align}
    \frac{\partial f}{\partial t} - \frac{\dot{B}}{B}\left( pP_2(\mu)\frac{\partial f}{\partial p} + \frac{3}{2}\mu(1 - \mu^2)\frac{\partial f}{\partial \mu} \right) = 0,
    \label{eq:DKE_spherical}
\end{align}
where $P_2(\mu)\equiv (3\mu^2 - 1)/2$ is the Legendre polynomial of order 2. 
By multiplying eqn. \ref{eq:DKE_spherical} by the particle energy $\epsilon$ and integrating over momentum space, we obtain 

\begin{align}
    \frac{d U}{dt} - \frac{\dot{B}}{B}\Delta P = 0,
    \label{eq:GyroviscousHeating}
\end{align}
where $\Delta P \equiv P_{\perp} - P_{\parallel}$ and 

\begin{align}
    P_{\perp}=\int \frac{p_{\perp}v_{\perp}}{2}f d^3p, & \ &
    P_{\parallel} = \int p_{\parallel}v_{\parallel}fd^3p,
\end{align}
which correspond to the pressures perpendicular and parallel to the magnetic field, respectively. The second term of the LHS in equation \ref{eq:GyroviscousHeating} corresponds to the gyroviscous heating rate for a pure shear driving motion, and it is valid for both nonrelativistic and relativistic particles (\cite{Kulsrud1983},\cite{ Hollweg1985}). Note that gyroviscosity can either heat or cool the plasma, depending on the relative signs of $\dot{B}/B$ and $\Delta P$; heating will proceed whenever they have the same sign and cooling whenever they have opposite signs. We will make use of this feature in \S \ref{sec:Results}.

We can also obtain an evolution equation for $\Delta P$ by noting that 
\begin{equation}
\label{eq:deltaPdef}
\Delta P = -\int pvP_2(\mu)fp^2dpd\mu. 
\end{equation}
Multiplying eqn. \ref{eq:DKE_spherical} by $-pvP_2(\mu)$, integrating over momentum space and assuming that the anisotropy remains small so the term $\partial f/\partial \mu$ can be dropped, we get

\begin{align}
    \frac{d\Delta P}{dt} - \frac{2}{5}\frac{\dot{B}}{B}\int \left( 4pv + p^2\frac{dv}{dp} \right)fp^2dp=0.
    \label{eq:AnisotropyEvolution}
\end{align}
Using $dv/dp=1/m\gamma^3$, the nonrelativistic and ultrarelativistic limits of eqn. (\ref{eq:AnisotropyEvolution}) are respectively

\begin{align}
    \frac{d\Delta P_{NR}}{dt} -2\frac{\dot{B}}{B} U = 0,
    \label{eq:AnisotropyEvolution_NR}
\end{align}


\begin{align}
    \frac{d\Delta P_R}{dt} -\frac{4}{5}\frac{\dot{B}}{B}U = 0.
    \label{eq:AnisotropyEvolution_R}
\end{align}
We can then write, in general,

\begin{align}
    \frac{d\Delta P}{dt} -\alpha \frac{\dot{B}}{B}U = 0,
    \label{eq:AnisotropyEvolution_gral}
\end{align}
where $4/5 < \alpha < 2$. Note that while eqn. (\ref{eq:AnisotropyEvolution_gral}) rests on the assumption of small anisotropy, eqn. (\ref{eq:GyroviscousHeating}) is exact.

Using equations \ref{eq:GyroviscousHeating} and \ref{eq:AnisotropyEvolution_gral} and changing the independent variable from $t$ to $\ln B$, we can see that there is a conserved quantity throughout the evolution,

\begin{align}
    Q \equiv \left(\frac{dU}{d\ln B}\right)^2 - \alpha U^2.
    \label{eq:ConservedQuantity}
\end{align}

Equation \ref{eq:ConservedQuantity} implies that, no matter how intricate its time evolution, whenever $B$ returns to its initial value, so does U. This way, we have shown that, in absence of a scattering mechanism, the evolution of $U$ is completely adiabatic irrespective of the evolution of $B$. 

The situation changes if we now allow the presence of scattering. To illustrate this, consider the addition of pitch angle scattering represented by a Lorentz operator so that equation \ref{eq:DKE_spherical} takes the form
\begin{align}
    \begin{split}
        \frac{\partial f}{\partial t} - \frac{\dot{B}}{B}\left( pP_2(\mu)\frac{\partial f}{\partial p} + \frac{3}{2}\mu(1 - \mu^2)\frac{\partial f}{\partial \mu} \right) \\ 
     =\frac{\partial }{\partial \mu}\left(\frac{\nu(1-\mu^2)}{2}\frac{\partial f}{\partial \mu}\right),
    \end{split}
    \label{eq:DriftKineticEquationScattering}
\end{align}
\begin{figure*}[t!]
    \centering
    \includegraphics[width=\textwidth]{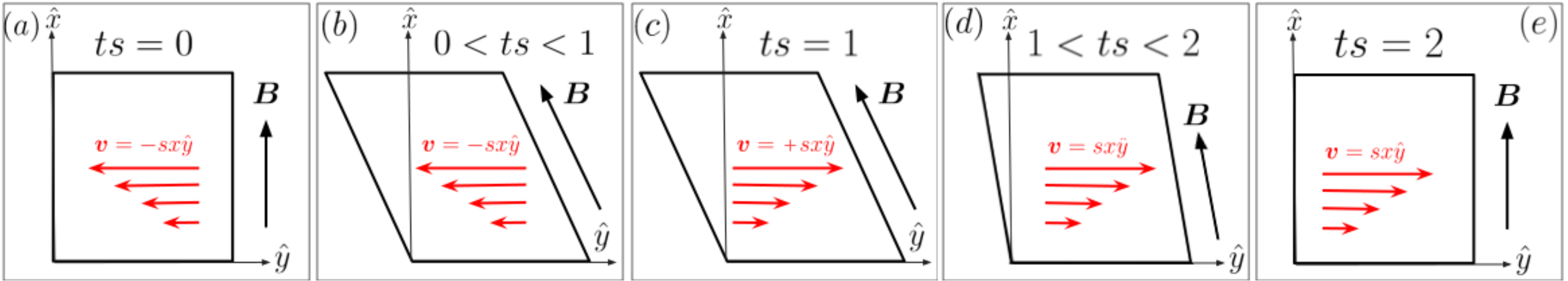}
    \caption{The simulation domain in our PIC simulations at different times. At $t\cdot s =0$ (panel $a$), the shear motion is applied (red arrows) and the domain follows the shearing flow of the plasma. Magnetic flux conservation changes the magnitude and orientation of the background magnetic field $\textbf{B}$, being amplified between $0<t\cdot s<1$ (panel $b$). At $t\cdot s = 1$ (panel $c$), the shear is reversed and the domain follows the shear back to the initial configuration. By magnetic flux conservation, the strength of the background magnetic field decreases in this phase until it reaches its initial value (panel $d$). At $t\cdot s = 2$ the domain is straight again and the shear motion continues and produces, again, an amplification of the magnetic field strength. The motion between $0<t\cdot s<2$ constitutes one pump cycle, and it repeats until the end of the simulation.}
    \label{fig:PeriodicShear}
\end{figure*}
where $\nu$ is the scattering rate which, for simplicity, we assume is independent of $\mu$. The source of scattering can be either Coulomb collisions, interactions with a spectrum of electromagnetic fluctuations originating from self-generated microinstabilities or just the high-frequency component of some pre-imposed turbulence.

While pitch angle scattering does not heat the plasma in and of itself, it does have an indirect effect through the evolution of the pressure anisotropy. Assuming small anisotropy as before, equation \ref{eq:AnisotropyEvolution_gral} is modified to 

\begin{align}
    \frac{d\Delta P}{dt} = \alpha\frac{\dot{B}}{B}U - 3\nu\Delta P,
    \label{eq:phaselag}
\end{align}
with the $\nu$ term resulting from breaking the adiabaticity of the evolution. The effect of this is most easily seen if we consider small amplitude, periodic changes in $B$, e.g. $\dot B/B=\epsilon\sin{\omega t}$, where $\epsilon\ll 1$, and solve eqn. (\ref{eq:phaselag}) to first order in $\epsilon$. Whereas for $\nu=0$, $\Delta P$ and $\dot B/B$ are out of phase by $\pi/2$ and yield no net heating when integrated over one magnetic cycle period, collisions introduce a phase shift in $\Delta P$ which is maximized for $\omega=\nu/3$,  irreversibly heating the plasma (see \cite{Lichko2017}, where $\nu$ is treated as a free parameter ), 
A more general model of this effect is presented in
\S\ref{sec:ScatteringModel}.

In what follows, we will study how gyroviscosity can effectively heat the plasma when the magnetic field varies periodically in time and there is a scattering source provided self consistently by kinetic microinstabilities such as mirror and firehose, in conditions similar to what can be encountered in the ICM (\cite{Borovsky1986,Schekochihin2005,Lyutikov2007}).

\section{Simulation Setup} 
\label{sec:Simulation Setup}

We use the fully kinetic, relativistic Particle in Cell (PIC) code TRISTAN-MP (\cite{Buneman1993}, \cite{Spitkovsky2005}) to simulate a shearing plasma made of singly charged ions and electrons (\cite{Riquelme2012}). The plasma has a homogeneous magnetic field that initially points along the $x-$axis, $\textbf{B}=B_0\hat{x}$. A periodic shear velocity field is imposed in our 2D domain (via shearing box coordinates; see \citep{Riquelme2012}), such that initially $\textbf{v} = -sx\hat{y}$ (red arrows in fig. \ref{fig:PeriodicShear}), where $x$ is the distance along the $x-$axis and $s$ is the shear rate. The velocity field then periodically changes its sign after a period of time $\tau_s = s^{-1}$, causing periodic reversals in the shear motion, as shown in fig. \ref{fig:PeriodicShear}. By magnetic flux conservation, the background magnetic field will vary accordingly, in both direction and magnitude. Initially, the shear amplifies the magnetic field $\textbf{B}$ in one direction during one shear time $\tau_s$, such that its $y-$component evolve as $dB_y/dt = -sB_0$, while $dB_x/dt=dB_z/dt=0$ (see figs. \ref{fig:PeriodicShear}$a$ and \ref{fig:PeriodicShear}$b$). After one shear time $\tau_s$, the shear velocity changes its sign and causes the magnetic field to decrease during another shear time $\tau_s$, such that now $dB_y/dt=sB_0$, while $dB_x/dt=dB_z/dt=0$, until it reaches its initial magnitude and direction (see figs. \ref{fig:PeriodicShear}$c$, \ref{fig:PeriodicShear}$d$ and \ref{fig:PeriodicShear}$e$). This way, in an interval equal to $2\tau_s$ a full cycle is completed, with $B$ having a maximum increase of $\sqrt{2}$ 
. Subsequently, a similar, ``mirroring" cycle occurs, but in this case $dB_y/dt= sB_0$ initially, and then switches back to $dB_y/dt= -sB_0$. This way, in an interval equal to $4\tau_s$, two cycles involving opposite shear motions in the y-direction are completed. The simulations cover several cycles, which is essential to assess the efficiency of the magnetic pumping mechanism in the secular heating of the plasma. 

In this configuration and in the absence of scattering, the particle magnetic moment $\mu_j=p^2_{\perp,j}/(2m_jB)$ and longitudinal action $\mathcal{J}_j = \oint p_{j,\parallel} d\ell$ are adiabatic invariants when $s\ll \omega_{c,j}$, where $\omega_{c,j}=eB/m_jc$ is the cyclotron frequency of particles of species $j$ and $e$ is the magnitude of the electric charge. This adiabatic invariance drives different pressure anisotropies ($P_{\perp,j}>P_{\parallel,j}$ and $P_{\perp,j}<P_{\parallel,j}$) depending on the variation of the magnetic field strength during the cycles, allowing--- once the anisotropy exceeds a marginal stability threshold --- the rapid excitation of kinetic instabilities that limit the anisotropy growth. 

Given the parameter space here explored, when $P_{\perp,j}>P_{\parallel,j}$, the mirror instability is excited (\cite{Chandrasekhar1958,RudakovSagdeev1961,Hasegawa1969,Southwood&Kivelson1993,Pokhotelov2002,Pokhotelov2004}). Mirror modes are nonpropagating, with wavevector highly oblique with respect to the mean magnetic field $\textbf{B}$ and maximum growth rate at $k_{\perp}R_{L,i}\sim 1$, where $R_{L,i}$ is the Larmor radius of the ions. It also presents Landau resonances with ions with very low parallel velocities with respect to $\textbf{B}$. 

On the other hand, when $P_{\perp,j}<P_{\parallel,j}$, the oblique firehose instability is excited (\cite{Yoon1993,Hellinger2000,Hellinger2008}). These modes are also nonpropagating, with wavevectors oblique with respect to $\textbf{B}$ with values of $kR_{L,i}$ slightly smaller than $\sim 1$. This instability is also resonant with ions through cyclotron resonances. In our simulations we self-consistently excite both mirror and oblique firehose instabilities throughout the entire evolution.

The physical parameters of the plasma are the initial temperature of the ions and electrons ($T_i$ and $T_e$), the initial ratio between the ion pressure $P_i^{init}$ and the magnetic pressure $\beta_{i}^{\text{init}}\equiv 8\pi P_i^{\text{init}}/B_0^2$, the mass ratio between ions and electrons $m_i/m_e$ and the ion ``magnetization'', defined as the ratio between the initial ion cyclotron frequency $\omega_{c,i}^{\text{init}}$ and the shear rate $s$, where $\omega_{c,i}^{\text{init}}=eB_0/m_ic$, with $B_0$ the initial magnetic field strength.

All the simulations reported here start with $\beta_{i}^{\text{init}}=20$ and $k_B T_i/m_i c^2 = 0.1$. Ions and electrons are initialized with Maxwell-Jüttner distributions (the relativistic generalization of Maxwell-Boltzmann) with $T_i=T_e$. A range of values for the mass ratio $m_i/m_e$ is used in our simulations and, given the current computational capabilities, using the realistic mass ratio $m_i/m_e\approx 1836$ becomes prohibitively expensive. The consequences of this constraint will be considered carefully and our findings indicate that the mass ratio does not play a significant role in the main results here reported. A range of values for the magnetization parameter $\omega_{c,i}^{\text{init}}/s$ is also used and the dependency of the results on this parameter will be discussed in the context of the turbulence expected in astrophysical environments such as the ICM. We note that the initial temperature of electrons in our simulations make them more relativistic than ions, and this limits the applicability of our results in the context of the ICM. 

The numerical parameters in our runs are the number of macroparticles per cell ($N_{\text{ppc}}$), the electron skin depth in terms of grid point spacing ($c/\sqrt{\omega_{p,e}^2+\omega_{p,i}^2}/\Delta x$, where $\omega_{p,e}^2=4\pi n_e e^2/m_e$ is the square of the electron plasma frequency and $n_e$ is the electron number density), and the box size in terms of the initial ion Larmor radius ($L/R_{L,i}^{\text{init}}$, where $R_{L,i}^{\text{init}}=v_{\text{th},i}/\omega_{c,i}^{\text{init}}$ and $v_{th,i}^2 = k_B T_i/m_i$). Table \ref{table:SimulationParameters} lists the names and parameters of our main simulations. We use $c/\sqrt{\omega_{p,e}^2+\omega_{p,i}^2}/\Delta x=3.5$ in all simulations listed in table \ref{table:SimulationParameters}. A series of simulations were performed to ensure that numerical convergence have been reached in every case in terms of $N_{\text{ppc}}, c/\omega_{p,e}/\Delta x$ and $L/R_{L,i}^{\text{init}}$. Those simulations are not shown in Table \ref{table:SimulationParameters}.

\begin{table}[t]
\centering
\caption{Parameters of the Simulations. $\beta_i^{\text{init}}\equiv 8\pi P_i^{\text{init}}/B_0^2$ is the initial ion plasma beta, $m_i/m_e$ is the ion to electron mass ratio, $\omega_{c,i}^{\text{init}}/s$ is the magnetization, where $\omega_{c,i}^{\text{init}} $ is the initial ion gyrofrequency and $s$ is the shear frequency, $N_{\text{ppc}}$ is the number of particles per cell, and $L/R_{L,i}^{\text{init}}$ is the size of the box in units of the initial ion Larmor radius. The reference simulation Zb20m2w800 is highlighted in bold and our characterization of numerical heating is based on the E simulations. }
\begin{tabular}{cccccc}
\hline \hline
Runs        & $\beta_i^{\rm init}$ & $m_i/m_e$ & $\omega_{c,i}/s$ & $N_{\rm{ppc}}$ & $L/R_{L,i}^{\rm{init}}$ \\ \hline
Zb20m2w200   & 20                   & 2         & 200              & 800            & 63                      \\
\textbf{Zb20m2w800}   & \textbf{20}                   & \textbf{2}         & \textbf{800}              & \textbf{800}             & \textbf{75}                      \\
Zb20m2w1600  & 20                   & 2         & 1600             & 520             & 75                      \\
Zb20m8w200   & 20                   & 8         & 200              & 40             & 55                      \\
Zb20m8w800   & 20                   & 8         & 800              & 40             & 55                      \\
Zb20m8w1600  & 20                   & 8         & 1600             & 40             & 55                      \\
Zb20m32w800  & 20                   & 32        & 800              & 40             & 51                      \\
Zb2m8w800 & 2                   & 8        & 800             & 400                  & 68                      \\
Eb20m2w200   & 20                   & 2         & 200              & 1600            & 63                      \\
Eb20m2w800   & 20                   & 2         & 800              & 800             & 63                      \\
Eb20m8w800   & 20                   & 8         & 800              & 40              & 55                      \\
\hline
\end{tabular}
\label{table:SimulationParameters}
\end{table}

\begin{figure*}[hbtp]
    \centering
    \includegraphics[width=\linewidth]{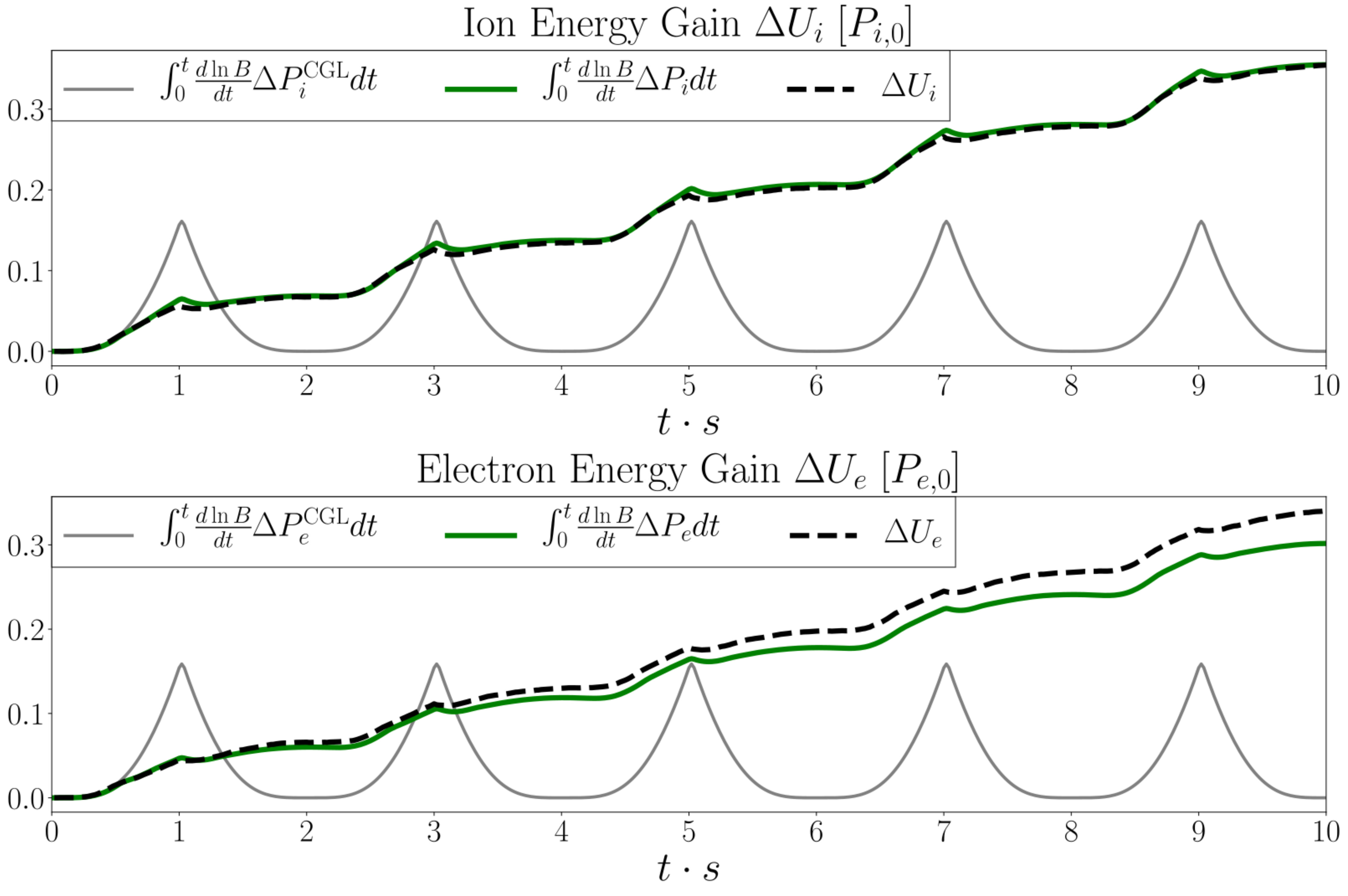}
    \caption{The effective ion and electron heating by gyroviscosity as a function of time for our reference simulation Zb20m2w800 (see table \ref{table:SimulationParameters}). All quantities are normalized by the initial pressure $P_{j,0}$ ($j=i,e$). The ion (upper panel) and electron (lower panel) energy gains $\Delta U_j = U_j(t)-U_j(0)$ are shown in dashed black lines after being corrected by numerical heating (see Appendix \ref{sec:NumericalHeating}). The heating by gyroviscosity integrated over time is shown in solid green lines (see eq. \ref{eq:GyroviscousHeating}). For comparison, the integrated heating by gyroviscosity considering a pure double-adiabatic CGL evolution for the pressure anisotropy $\Delta P_j$ \textbf{(i.e. without scattering)} is shown in solid gray lines. The units on this plot and all subsequent heating plots are such that the ordinate represents the temperature increment in units of initial temperature}
    \label{fig:gyroviscousheating}
\end{figure*}

In our simulations, both ions and electrons exhibit an excess energy gain due to numerical heating related to the inherent presence of electric field fluctuations generated by the discreteness in the number of particles in PIC simulations. We studied how this numerical heating behaves in appendix \ref{sec:NumericalHeating} by performing additional simulations designed to isolate the effect of numerical heating. We then subtracted it from the ion and electron internal energy $U_i(t), U_e(t)$ here presented. This allowed us to reduce the discrepancy in energy gain to $\sim 0.1\%$ for ions and $\sim 12\%$ for electrons (see fig. \ref{fig:gyroviscousheating}). While subtracting the numerical heating rate does not completely nullify the effect of numerical heating because the gyroviscous heating rate is probably not exactly the same as if numerical heating were not present, we hope it does indicate the size of this numerical effect.


\section{Results}
\label{sec:Results}

In this section we present our results of the periodic shear simulations described in section \S \ref{sec:Simulation Setup}. We show that heating by gyroviscosity (cf. eq. \ref{eq:GyroviscousHeating}) can effectively be retained after one pump cycle, with an efficiency that depends on the evolution of the mirror and firehose instabilities in each phase of the cycle. We also discuss how this depends on physical parameters like the mass ratio $m_i/m_e$ and magnetization $\omega_{c,i}^{\text{init}}/s$.

\subsection{Heating by Gyroviscosity}
\label{subsec:GyroviscosityHeating}

Figure \ref{fig:gyroviscousheating} shows the evolution of the internal energy gain $\Delta U_j \equiv U_j(t) - U_j(0)$ after being corrected for numerical heating in dashed black line and of the integrated gyroviscous heating rate in solid green line (cf. eq.  \ref{eq:GyroviscousHeating}) as a function of time for run Zb20m2w800. Importantly, we have found that, over the 5 cycles shown, the electron and ion temperatures have each increased by about 30\%.

We can see that both ions and electrons gain energy in each pump cycle, and the energy evolution is well tracked by the action of gyroviscosity. The energy gain is irreversible and this becomes apparent by comparing it with the expected integrated heating rate if $\Delta P_j$ evolved in the absence of any pitch-angle scattering, so the adiabatic invariants of the particles remain constant: $\Delta P_{j,\text{CGL}}=P_{\perp,\text{CGL}}-P_{\parallel,\text{CGL}} = P_{\perp,0}B/B_0 - P_{\parallel,0}(B_0/B)^2$ (i.e., a double-adiabatic evolution, \cite{CGL1956}, solid gray line). We can see that, indeed, when the particles' adiabatic invariants are not broken, the heating by gyroviscosity in one pump cycle is completely reversible, as predicted in section \ref{sec:TheoreticalModel} (see eq. \ref{eq:ConservedQuantity}). Consequently, the breaking of the adiabatic invariants of the particles by the presence of pitch-angle scattering or other isotropizing processes in our simulations is what allows the plasma to redistribute the energy between perpendicular and parallel components and effectively retaining part of the energy after a pump cycle.
By looking at the first pump cycle ($0<t\cdot s<2$), we can see that at $t\cdot s =1$ the integrated gyroviscous heating is smaller than the expected heating in absence of scattering. This, however, is followed by a much shorter period of gyroviscous cooling ($1<t\cdot s<1.3$), and then followed by another period of gyroviscous heating ($1.3<t\cdot s<2$). The smaller heating obtained in the first cycle is then compensated with a much shorter period of gyroviscous cooling that allows most of the heat to be retained at the end of the cycle. This behavior is directly related to the evolution of the pressure anisotropy and $\dot{B}/B$ during the pump cycle. This evolution is determined by the excitation of mirror and firehose instabilities, which are the agents that provide the pitch-angle scattering in our simulations, as we will show in \S \ref{sec:AnisotropyEvolution}. Note that although gyroviscous heating accounts for the ion energy gain almost entirely, there is a small additional amount of electron heating. This might be because the wave-particle interactions are not completely elastic (see discussion in section \ref{sec:Anisotropy_mime_wcis_Comparison}) or because numerical heating acts differently on the electrons. However, we have not pinpointed the cause. 

\subsection{Mass Ratio and Magnetization Dependence}

\begin{figure}[hbtp]
    \centering
    \begin{tabular}{c}
         \includegraphics[width=\linewidth]{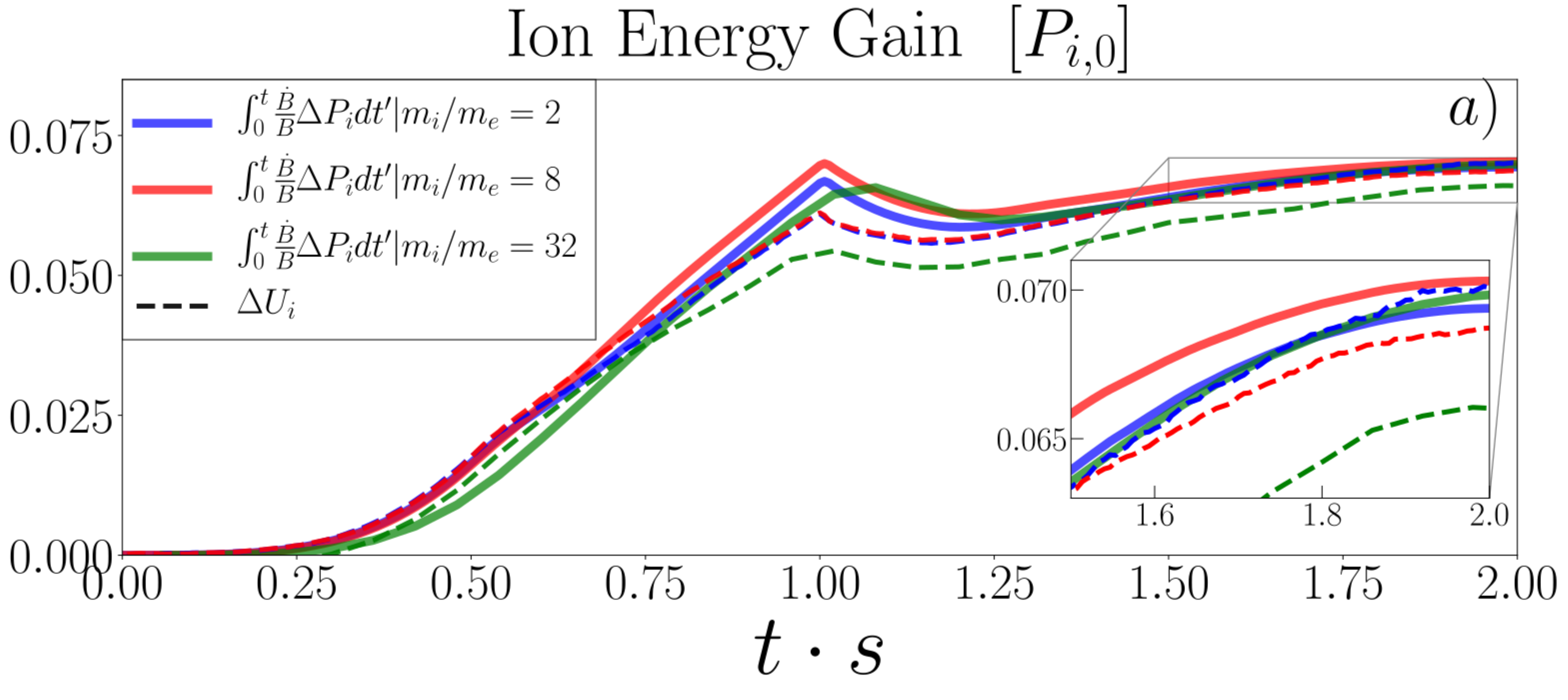}\\
         \includegraphics[width=\linewidth]{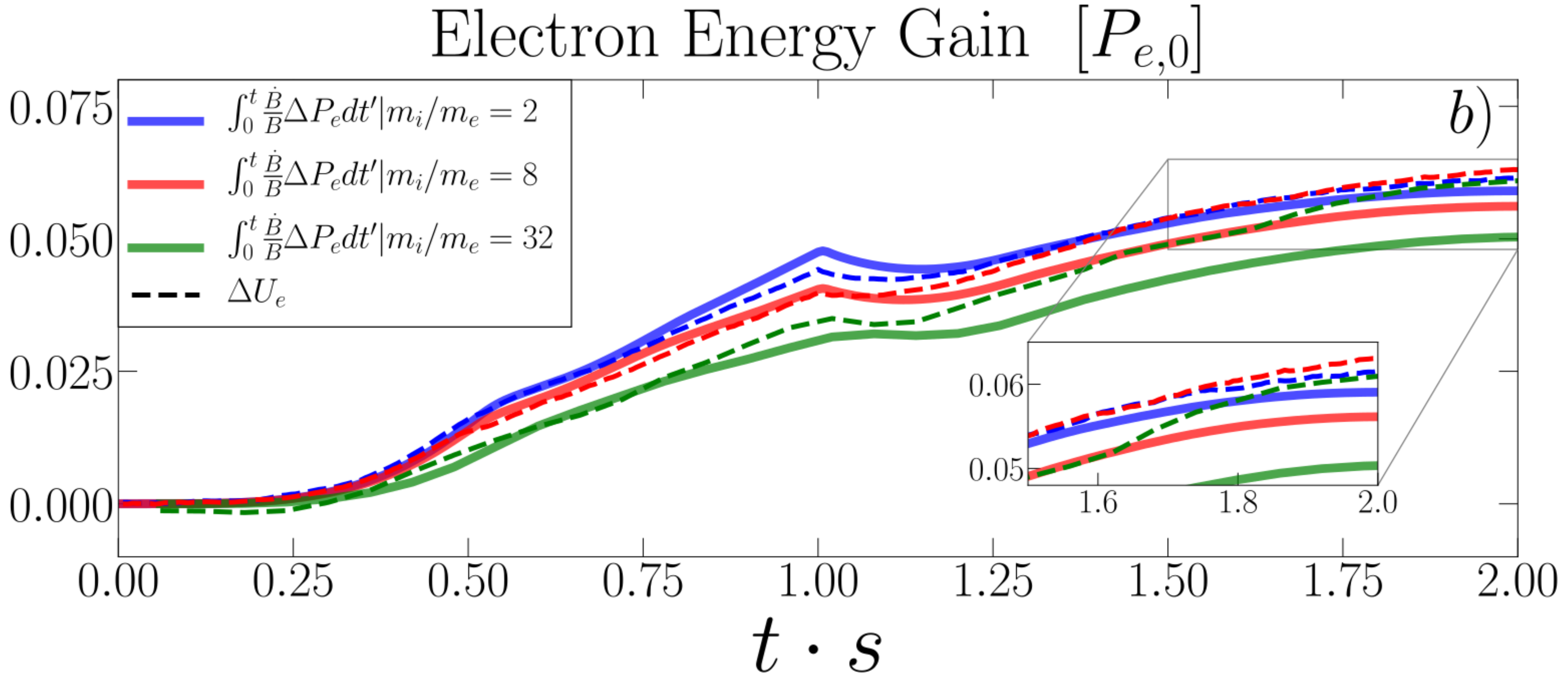} \\
         \includegraphics[width=\linewidth]{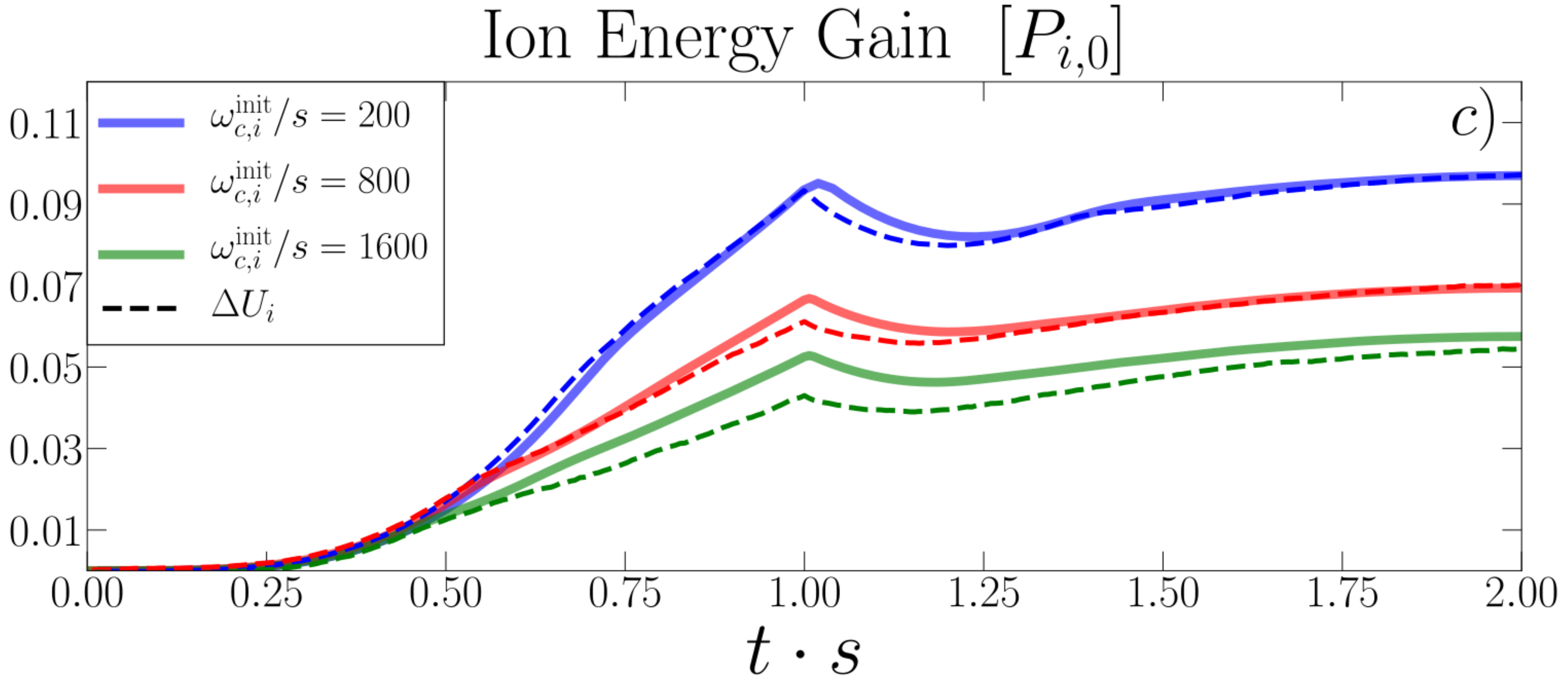} \\
         \includegraphics[width=\linewidth]{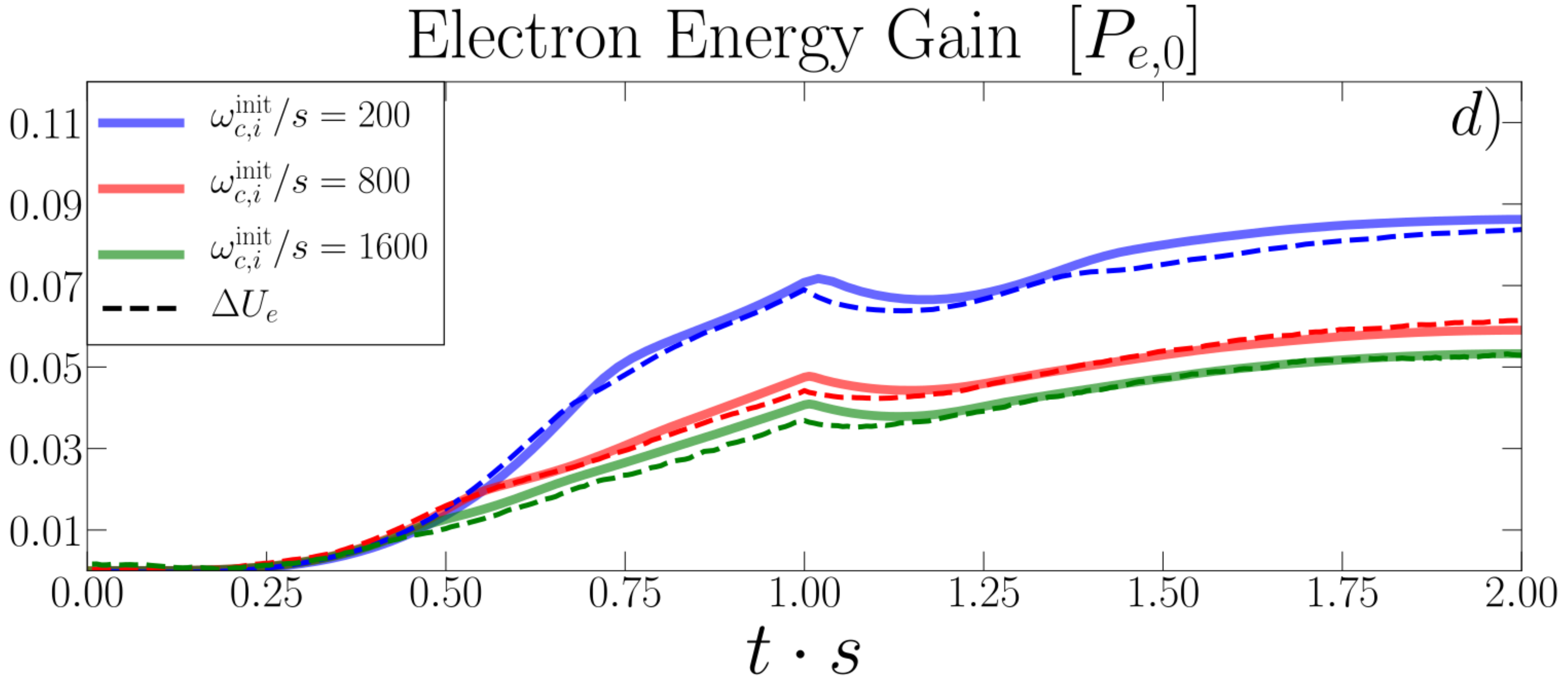}
    \end{tabular}
    \caption{Heating dependency on mass ratio $m_i/m_e$ and magnetization $\omega_{c,i}^{\text{init}}/s$. Panel $(a)$: the dashed lines show the evolution of the ion energy gain $\Delta U_i = U_i(t)-U_i(0)$ after being corrected by numerical heating and the solid lines show the expected integrated rate of heating by gyroviscosity $\int_0^t dt' \Delta P_i d\ln B/dt $ for runs Zb20m2w800, Zb20m8w800 and Zb20m32w800, all with same magnetization $\omega_{c,i}^{\text{init}}/s=800$ and mass ratios of $m_i/m_e=2$ (blue lines), $m_i/m_e=8$ (red lines) and $m_i/m_e=32$ (dashed green line), respectively. Panel $(b)$: same as panel $(a)$ but for electrons. Panel $(c)$: the evolution of the ion energy gain $\Delta U_i = U_i(t)-U_i(0)$ (dashed lines) and the expected integrated heating rate by gyroviscosity $\int_0^t \dot{B}\Delta P_i/B dt'$ (solid lines) for runs Zb20m2w200, Zb20m2w800 and Zb20m2w1600, all with same mass ratio $m_i/m_e=2$ and magnetizations of $\omega_{c,i}^{\text{init}}/s=200$ (blue line), $\omega_{c,i}^{\text{init}}/s=800$ (red line) and $\omega_{c,i}^{\text{init}}/s=1600$ (green line), respectively. Panel $(d)$: same as panel $(c)$ but for electrons}
    \label{fig:HeatingDependency}
\end{figure}

Figure \ref{fig:HeatingDependency}$a$ shows the evolution of the ion energy gain $\Delta U_i = U_i(t)-U_i(0)$ in dashed lines after being corrected by numerical heating (see appendix \ref{sec:NumericalHeating}) and the evolution of the integrated gyroviscous heating rate in solid lines for runs Zb20m2w800, Zb20m8w800 and Zb20m32w800, with same magnetization $\omega_{c,i}^{\text{init}}/s=800$ and mass ratios of $m_i/m_e=2$, $m_i/m_e=8$ and $m_i/m_e=32$, respectively. We can see that there is a net heating in all three cases, the energy gain is well tracked by the gyroviscous heating, and it does not exhibit a significant dependence on mass ratio. A similar behavior is exhibited in the case of electrons in fig. \ref{fig:HeatingDependency}$b$, where a net heating is also obtained and the mass ratio does not play a significant role in their final energy gain either, although the difference is larger than in the case of ions.

Analogously, fig. \ref{fig:HeatingDependency}$c$ and \ref{fig:HeatingDependency}$d$ shows the evolution of the ion and electron energy gain $\Delta U_j = U_j(t)-U_j(0)$ $(j=i,e)$ in dashed lines after being corrected by numerical heating and the integrated gyroviscous heating rate in solid lines for runs Zb20m2w200, Zb20m2w800 and Zb20m2w1600, with same mass ratio $m_i/m_e=2$ and magnetizations $\omega_{c,i}^{\text{init}}/s=200$, $\omega_{c,i}^{\text{init}}/s=800$ and $\omega_{c,i}^{\text{init}}/s=1600$, respectively. We can see that in all cases a net heating is obtained after a pump cycle, and in this case the heating rate  decreases for larger magnetizations. This decrease in the heating rate is directly related to the evolution of the pressure anisotropy and the excitation of mirror and firehose instabilities, which depend on the magnetization parameter, as we will show in section \ref{sec:AnisotropyEvolution}.


Before proceeding further, we calculate the gyroviscous heating predicted by the
model of \cite{Kunz2011}\footnote{In \cite{Kunz2011} (e.g. eqn. (8)), the heating rate is written in terms of the stress tensor $\boldsymbol{\sigma}$, which can be shown to be equivalent to eqn. (\ref{eq:GyroviscousHeating}).}. In this model it is assumed that efficient scattering maintains $\Delta P$ at its marginally stable value
\begin{equation}\label{eq:k20111}
\Delta P = \frac{\xi_{\pm}}{\beta} P,
\end{equation}
where the $\xi_{\pm}$ are constants of order unity and have the same sign as $\dot B/B$,
with the plus and minus signs referring to increasing and decreasing $B$ respectively. Equation (\ref{eq:GyroviscousHeating}) can then be written in the form
\begin{equation}\label{eq:k20112}
\frac{dU}{dt}=  \frac{\xi_{\pm}}{16\pi}\frac{dB^2}{dt}.  
\end{equation}
For our magnetic field model, $B^2$ doubles during the shearing phase of the cycle (i.e. $0<t\cdot s<1$) and returns to its original value in the deshearing phase ($1<t\cdot s<2$). Equation (\ref{eq:k20112}) can then be integrated over $0\le t\cdot s\le 1$ and then over $1\le t\cdot s\le 2$ to give the temperature change $\Delta T$ over a complete cycle. Assuming a nonrelativistic equation of state, we find 
\begin{equation}\label{eq:k20113}
\frac{\Delta T}{T_0}  =\frac{\xi_+ - \xi_-}{3\beta_0}. 
\end{equation}
Choosing $\xi_+=1$, $\xi_{-}=-1.4$ and $\beta_0 = 40$ (see \S \ref{sec:PressureAnisotropyMirrorFirehose}), the marginal stability model predicts an energy gain of about 2\% per cycle, which is significantly less than the heating rate shown in Fig. \ref{fig:gyroviscousheating}. In the next section, we show that the enhanced heating rate occurs because $\Delta P$ exceeds its marginal stability value for the high frequency turbulence
considered here.

\section{Anisotropy Evolution and Instability Excitation}
\label{sec:AnisotropyEvolution}
Here we describe the evolution of the pressure anisotropy $\Delta P_j$ in our simulations. After a brief linear phase consistent with CGL double-adiabatic scaling (\cite{CGL1956}), the nonlinear evolution of $\Delta P$ is determined by the alternating excitation of the mirror and firehose instabilities during a pump cycle, limiting the anisotropy growth and providing the necessary scattering for the operation of magnetic pumping. It is this interplay which sets the efficiency of the gyroviscous heating in a pump cycle.

\subsection{Pressure Anisotropy and Mirror/Firehose Evolution}
\label{sec:PressureAnisotropyMirrorFirehose}
\begin{figure*}[t]
    \centering
    \includegraphics[width=\linewidth]{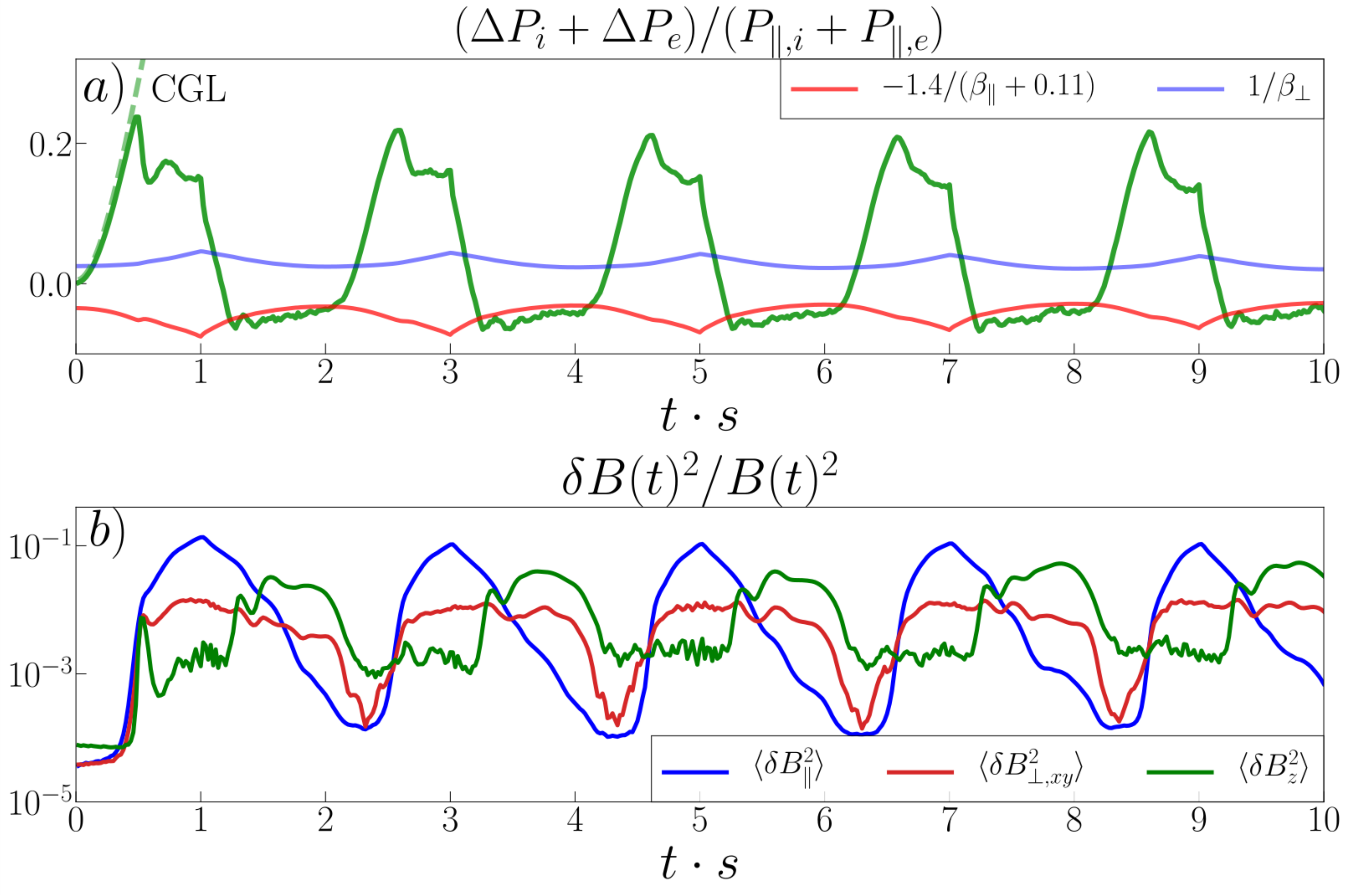}
    \caption{Panel $(a)$: The evolution of the total pressure anisotropy $\Delta P \equiv \Delta P_i+\Delta P_e$ for the reference simulation Zb20m2w800 in table \ref{table:SimulationParameters} (green line). The dashed green line shows the corresponding double-adiabatic prediction (CGL,\cite{CGL1956}). As a reference, the solid red and blue lines show approximate linear thresholds for the excitation of the mirror and firehose instabilities, respectively, where $\beta_{\perp}\equiv \beta_{i,\perp}+\beta_{e,\perp}$ and $\beta_{\parallel}\equiv \beta_{i,\parallel}+\beta_{e,\parallel}$ (\cite{Hasegawa1969,Hellinger2008}). Panel $(b)$: volume-averaged magnetic energy in $\delta \textbf{B}$ along different axes and as a function of time for run Zb20m2w800. Here $\delta B_{\parallel}$ (blue line) is the component parallel to $\langle \textbf{B} \rangle$, $\delta B_{\perp,xy}$ (red line) and $\delta B_z$ (green line) are, respectively, the components perpendicular to $\langle \textbf{B} \rangle$ in the plane and perpendicular to the plane of the simulation.}
    \label{fig:DeltaP}
\end{figure*}

\begin{figure}[t]
    \centering
    \begin{tabular}{c}
        \includegraphics[width=\linewidth]{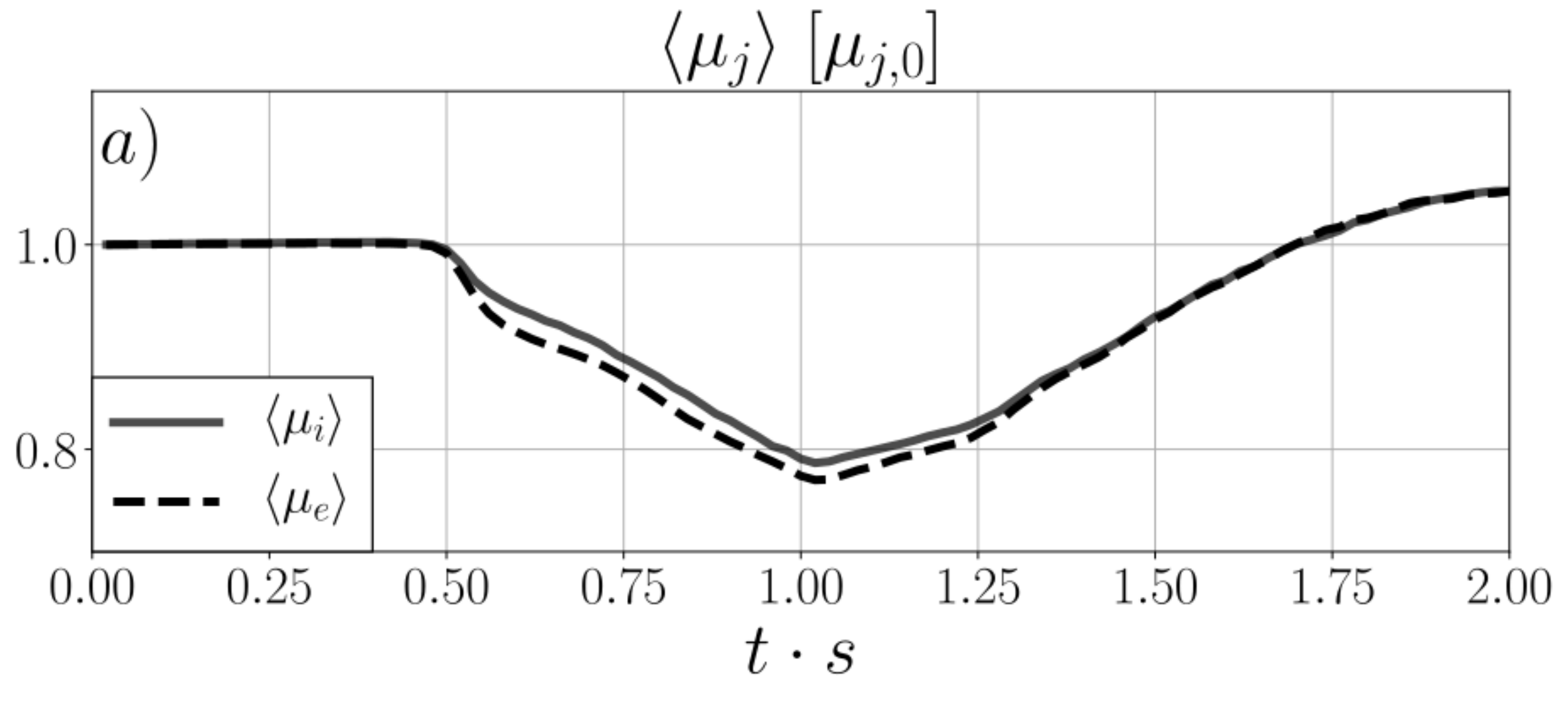}   \\
        \includegraphics[width=\linewidth]{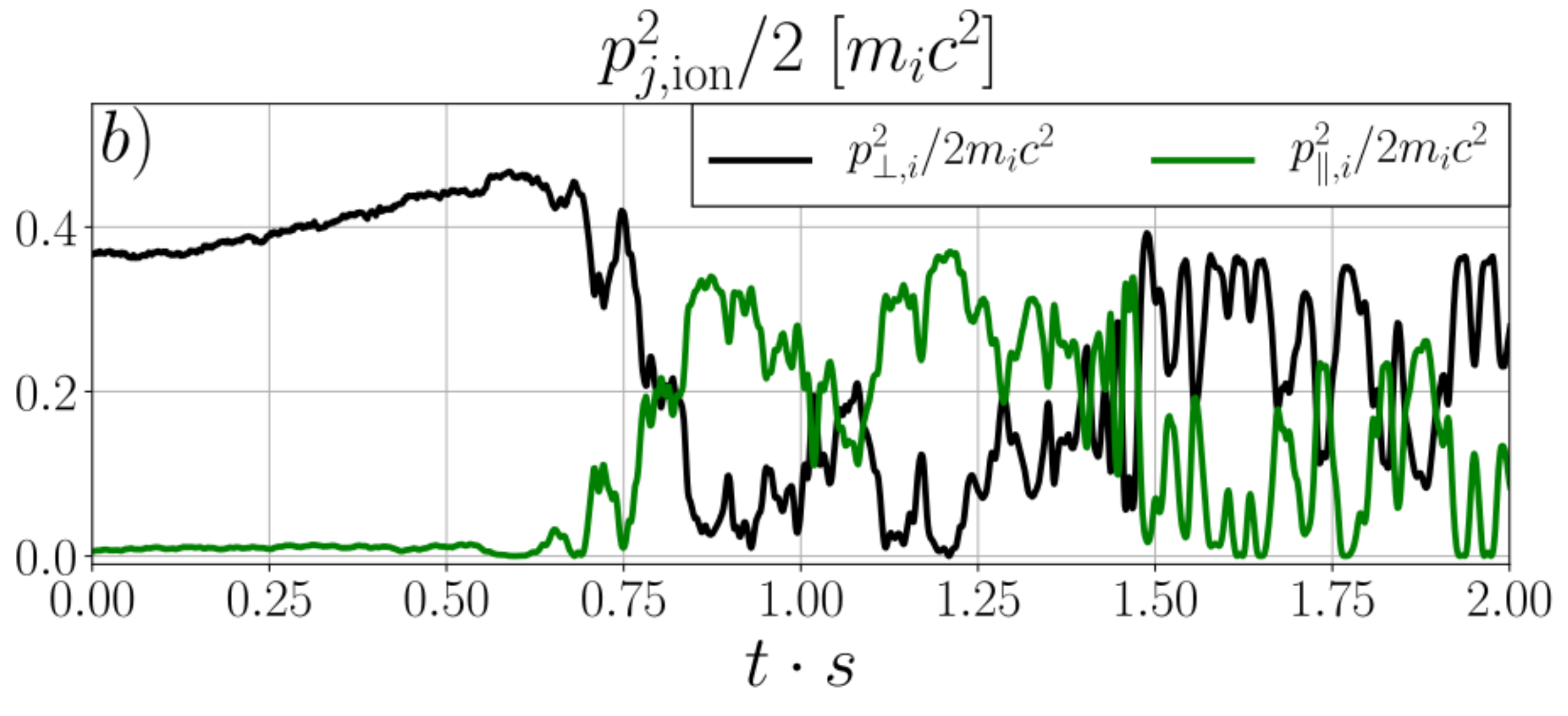}  
    \end{tabular}
    \caption{Panel $a$: The evolution of the volume-averaged ion (solid line) and electron (dashed line) magnetic moments $\langle \mu_j \rangle = \langle p_{\perp,j}^2/B \rangle (j=i,e)$ in one pump cycle for run Zb20m2w800. Panel $b$: The evolution of the perpendicular (solid black line) and parallel (solid green line) kinetic energy of one ion in one pump cycle for run Zb20m2w800.}
    \label{fig:Magmom}
\end{figure}

\begin{figure}[t]
    \centering
    \includegraphics[width=\linewidth]{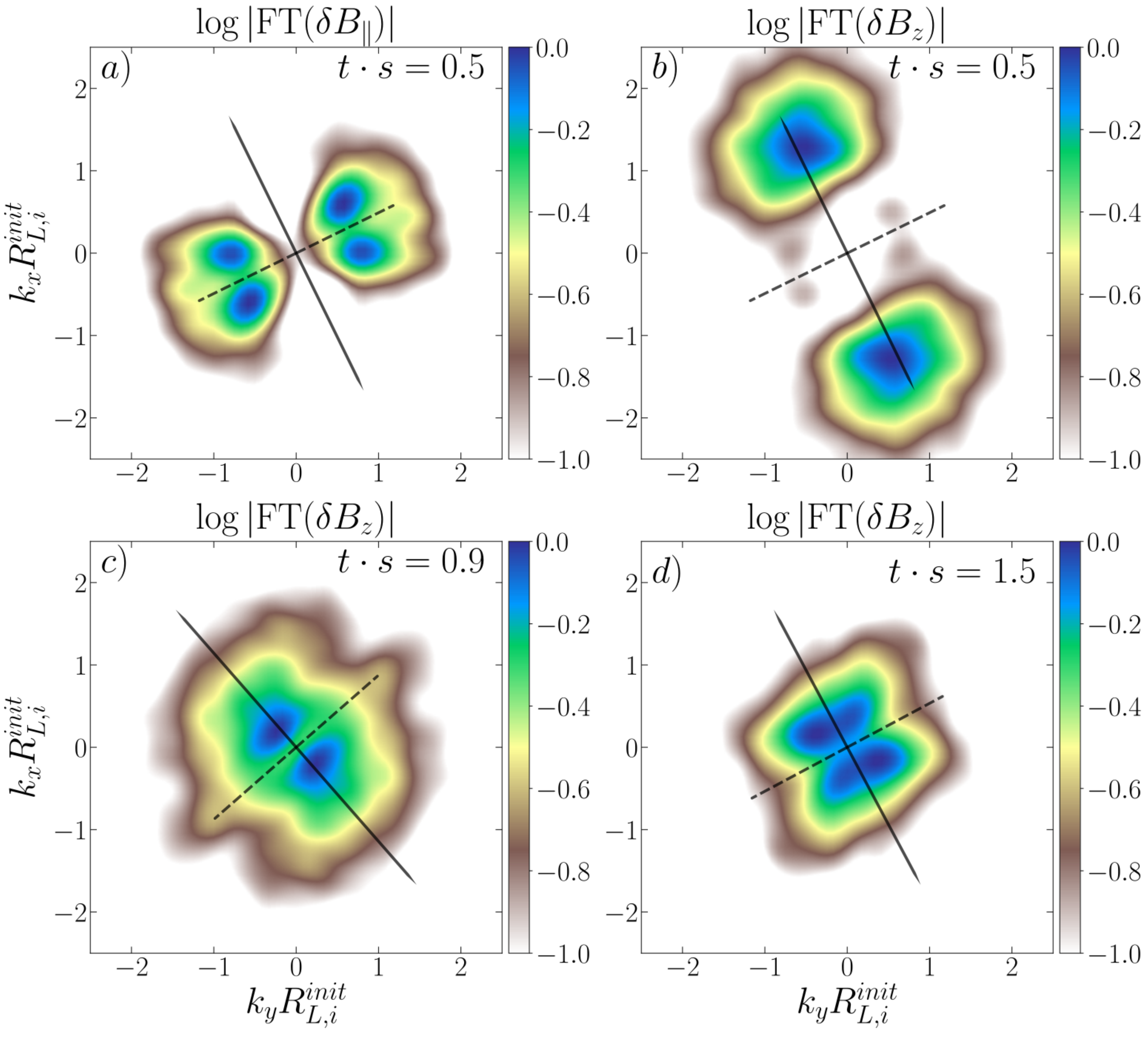}
    \caption{Fourier transform (FT) in space of two different components of $\delta\textbf{B}$ at four different times for run Zb20m2w800. The wavenumbers $k_x,k_y$ are normalized to the initial ion Larmor radius $R_{L,i}^{\text{init}}$. In all panels, the solid and dashed black lines represent, respectively, the direction along $\langle \textbf{B} \rangle$ and perpendicular to $\langle \textbf{B} \rangle$ at the corresponding time. Panel $(a)$: magnitude of the FT of $\delta B_{\parallel}$ at $t\cdot s=0.5$. Panel $(b)$: magnitude of the FT of $\delta B_z$ at $t\cdot s=0.5$. Panel $(c)$: magnitude of the FT of $\delta B_z$ at $t\cdot s=0.9$. Panel $(d)$: magnitude of the FT of $\delta B_z$ at $t\cdot s=1.5$.}
    \label{fig:BFluctuationsFFT_space}
\end{figure}

We can see the evolution of the pressure anisotropy and the development of mirror and firehose instabilities in fig. \ref{fig:DeltaP}. Figure \ref{fig:DeltaP}$a$ shows the total pressure anisotropy, $\Delta P \equiv \Delta P_i + \Delta P_e$ normalized by the total parallel pressure $P_{\parallel}\equiv P_{\parallel,i}+P_{\parallel,e}$ as a function of time in green solid line and fig. \ref{fig:DeltaP}$b$ shows the time evolution of the magnetic energy in $\delta \textbf{B}$ along different directions for run Zb20m2w800 in table \ref{table:SimulationParameters} ($\delta \textbf{B} \equiv \textbf{B} - \langle\textbf{B}\rangle$ and $\langle\rangle$ denotes a volume average over the entire simulation box). The decomposition of $\delta \textbf{B}$ is done in terms of the direction parallel to $\langle \textbf{B} \rangle$ ($\delta B_{\parallel}$, blue line), perpendicular to $\langle \textbf{B} \rangle$ in the plane of the simulation ($\delta B_{\perp,xy}$, red line) and perpendicular to $\langle \textbf{B} \rangle$ and out of the simulation plane ($\delta B_z$, green line).  Initially, the particle distribution is isotropic. The initial amplification of the magnetic field drives $\Delta P > 0$ and an early evolution consistent with double-adiabatic scalings (dashed green line). The mirror instability's hydromagnetic threshold $1/\beta_{\perp}$, where $\beta_{\perp} \equiv \beta_{i,\perp} + \beta_{e,\perp}$ (\cite{Hasegawa1969}), is shown as a solid blue line in fig. \ref{fig:DeltaP}$a$
\footnote{Even when the fastest growing mirror modes in our simulations have $kR_{L,i}\sim 1$ where $R_{L,i}$ is the Larmor radius of the ions (see fig. \ref{fig:BFluctuationsFFT_space}), we do not see important differences when comparing the evolution of $\Delta P$ with more accurate kinetic instability thresholds such as ones provided by \cite{Pokhotelov2004}; the anisotropy $\Delta P$ always largely surpasses the threshold for the excitation of the mirror instability and it does not stay at marginal stability afterwards.} .
At $t\cdot s \approx 0.5$ the ion and electron magnetic moments are broken by the excitation of mirror modes (see fig. \ref{fig:Magmom}$a$), when they start to interact with the particles. This way, the growth of $\Delta P$ is limited throughout the initial half-cycle by mirror modes. It is important to note that after surpassing the mirror instability threshold, $\Delta P$ does not reach a marginally stable state $\sim 1/\beta_{\perp}$, but it saturates at a larger value, $\Delta P/P_{\parallel} \sim 0.15$, similar to that reported by \cite{Melville2016}. We will come back to this point in \S\ref{sec:Anisotropy_mime_wcis_Comparison} where we compare different magnetization values. Note that the relative amplitudes of all components of $\delta\mathbf{B}$ are quite large, suggesting that linear wave physics alone does not describe these fluctuations.

\begin{figure*}[t]
    \centering
    \includegraphics[scale=0.32]{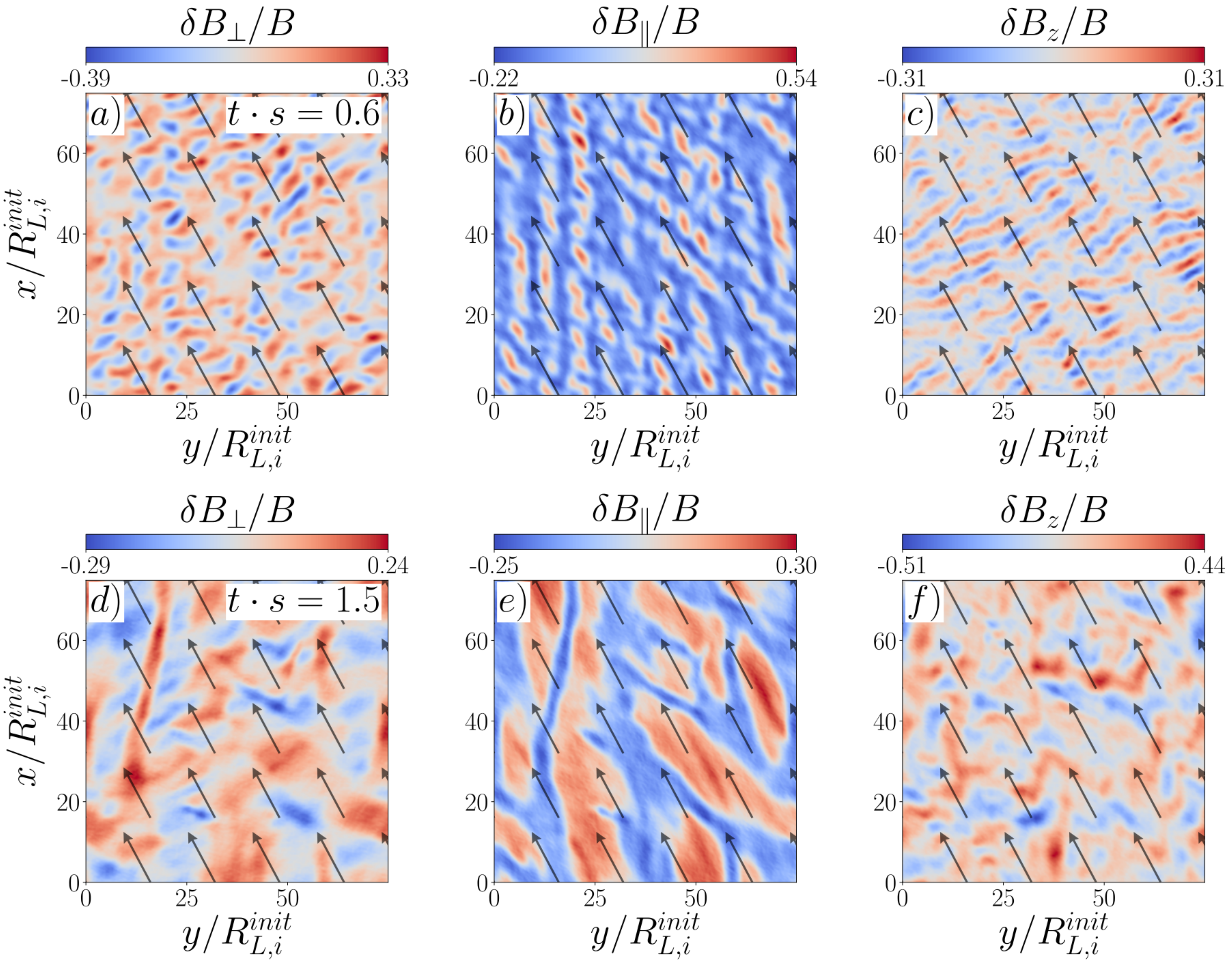}
    \caption{Upper row: Parallel and perpendicular components of the magnetic fluctuation $\delta \textbf{B}$ with respect to $\langle \textbf{B} \rangle$ for run Zb20m2w800 at time $t\cdot s=0.6$; left panel shows the in plane perpendicular component $\delta B_{\perp}$, middle panel shows the parallel component $\delta B_{\parallel}$ and right panel shows the out-of-plane perpendicular component, $\delta B_z$. The black arrows show the direction of the mean magnetic field $\langle \textbf{B}\rangle$ at $t\cdot s=0.6$. Bottom row: same as upper row but at time $t\cdot s = 1.5$. See suplemental material for an animated version of this figure over one pump cycle.}
    \label{fig:BFluctuations}
\end{figure*}

The excitation of the mirror instability is also consistent with the dominance of $\delta B_{\parallel}^2$ during the initial half-cycle, especially from $0.5 \lesssim t\cdot s \lesssim 1$. 
The growth of $\delta B_{\parallel}^2$ is accompanied by a similar growth of $\delta B_z^2$ that is quickly damped around $t\cdot s\approx 0.6$. Two explanations may be proposed for the origin of this feature in $\delta B_z^2$. On the one hand, it is consistent with the appearance of the non-coplanar component of mirror modes when the fastest growing mode has $kR_{L,i}\sim 1$ in the kinetic regime (\cite{Pokhotelov2004}, see fig. \ref{fig:BFluctuationsFFT_space}$a$ and \ref{fig:BFluctuationsFFT_space}$b$). On the other hand, the structure of the modes in $\delta B_z^2$ and their parallel propagation (see fig. \ref{fig:BFluctuations}$c$ and \ref{fig:BFluctuationsFFT_space}$b$) also resemble ion-cyclotron (IC) modes, although they are expected to be subdominant for large plasma $\beta$, based on linear theory and simulations (\cite{Riquelme2015}). The disentanglement of the true nature of these fluctuations has been reported to be a difficult task (\cite{Pokhotelov2004}). As the overall dynamics is still dominated by mirror modes at this stage of the pump cycle, the proper identification of the modes seen in $\delta B_z^2$ is inessential to this work. However, it is interesting to note that after its initial decay at $t\cdot s\approx 0.6$, $\delta B_z^2$ grows again and begins oscillating; $dB_{\perp,xy}^2$ begins to oscillate, too. We will see that these behaviors are consistent with the appearance of short-lived, parallel propagating whistler modes localized in the regions of low magnetic field associated with mirror modes, commonly known as ``lion roars''  (\cite{Baumjohann1999,Breuillard2018}), which will be investigated more completely in future work (Ley et al. 2022, in prep.).

At $t\cdot s = 1$, the shear is reversed and the magnetic field starts to dwindle, driving a decrease in $\Delta P$. Since $\dot B$ becomes negative while $\Delta P$ is positive, the plasma undergoes gyroviscous \textit{cooling}, but as shown in Fig. \ref{fig:gyroviscousheating} and \ref{fig:HeatingDependency}, the duration of this phase is short and the cooling is slight.  Because mirror instability has kept $\Delta P$ at a lower value than the one predicted by CGL, the decrease of $\Delta P$ does not end at a complete isotropic state at the end of the cycle, but grows further to $\Delta P < 0$, driving the system anisotropic again (but with opposite sign), and setting the conditions for the excitation of the oblique firehose instability. The approximate threshold $-1.4/(\beta_{\parallel}+0.11)$ for this instability is shown in solid red line in fig. \ref{fig:DeltaP}$a$, where $\beta_{\parallel}\equiv\beta_{i,\parallel}+\beta_{e,\parallel}$ (\cite{Hellinger2008}). Indeed, at $t\cdot s \approx 1.5$, we can see in fig. \ref{fig:DeltaP}$b$ that $\delta B_z^2$ is now the dominant component, consistent with firehose modes (\cite{Hellinger2000}). 

The pitch-angle scattering that these modes provide quickly stops $\Delta P$ from growing more negative, bringing it closer to isotropy until the end of the second half of the pump cycle ($t\cdot s=2$). In this case $\Delta P$ is able to reach values much closer to marginal stability and is well tracked by the oblique firehose instability threshold. The presence of firehose modes is also reflected in the evolution of the particle magnetic moment shown in fig. \ref{fig:Magmom}$b$. 

The scattering the particles are subjected to by waves is such that they exchange energy between parallel and perpendicular components in each interaction, and this can be clearly seen in fig. \ref{fig:Magmom}$b$, where the perpendicular and parallel kinetic energy (with respect to $\textbf{B}$) is shown for one ion during one pump cycle. During the first half of the pump cycle where mirror modes are developed ($t\cdot s \sim 0.6$), we see that there is a transfer of energy from perpendicular to parallel components, consistent with a reduction of the perpendicular pressure $P_{\perp}$. On the other hand, during the second half of the cycle when oblique firehose modes dominate ($t\cdot s\sim 1.5$) we see energy transfer in the opposite sense, from parallel to perpendicular components, consistent now with a reduction of the parallel pressure $P_{\parallel}$. This process is essential for retaining part of gyroviscous heating after a pump cycle (see section \ref{subsec:GyroviscosityHeating}). The next cycles present mainly the same phenomenology. 

This way, fig. \ref{fig:DeltaP}$a$ shows how the pressure anisotropy undergoes periodic stages of growth, decrease and negative growth, always bounded between values determined by the efficiency of the isotropizing processes provided by mirror and firehose modes (and not necessarily constrained by marginal stability). As we will show in section \ref{sec:Anisotropy_mime_wcis_Comparison}, this efficiency will depend on the magnetization parameter, $\omega_{c,i}^{\text{init}}/s$.

The overall alternating excitation of mirror and firehose instabilities in the parameter space studied here is qualitatively consistent with the results by \cite{Melville2016} for both mirror-to-firehose and firehose-to-mirror regimes. 
The presence of mirror and firehose modes at different stages of the pump cycle can be seen in fig. \ref{fig:BFluctuations}, where we show two snapshots of $\delta \textbf{B}$ in the same three different directions used in fig. \ref{fig:DeltaP}$b$, and black arrows representing the direction of $\langle \textbf{B}\rangle$. Because $\langle\mathbf{B}\rangle$ is almost the same in the two cases, the clear differences in the character of the fluctuations can only be attributed to the anisotropy of the plasma. 

In fig. \ref{fig:BFluctuations}$b$ ($t\cdot s = 0.6$), we can see that $\delta B_{\parallel}$ has the largest amplitude, and the structure of oblique mirror modes can also be observed. This is consistent with the expectation for mirror modes having their components mainly in the plane $(\textbf{k},\textbf{B})$ (\cite{Pokhotelov2004}). On the other hand, in fig. \ref{fig:BFluctuations}$f$ ($t\cdot s = 1.5$), we can see that now $\delta B_z$ is the largest component, exhibiting the presence of oblique firehose modes, consistent with the dominance of $\delta B_z^2$ at this time (see fig. \ref{fig:DeltaP}$b$).

The oblique nature of mirror modes is also evident from fig. \ref{fig:BFluctuationsFFT_space}$a$. The spatial Fourier transform of $\delta B_{\parallel}$ is shown at $t\cdot s =0.5$, with the solid and dashed black lines showing the direction along and perpendicular to $\langle \textbf{B} \rangle$ at $t\cdot s=0.5$, respectively. The modes span angles $\theta_k = \cos^{-1}(\textbf{k}\cdot \langle\textbf{B}\rangle/kB)$ between $\sim 45^{\circ}$ and $\sim 90^{\circ}$, although most of the power is concentrated in a range $60^{\circ}\lesssim\theta_k\lesssim 80^{\circ}$. 

We can better characterize the two different stages seen in the evolution of $\delta B_z^2$ during $0\lesssim t\cdot s\lesssim 0.6$ and $0.6\lesssim t\cdot s\lesssim 1.2$ in panels \ref{fig:BFluctuationsFFT_space}$b$ and \ref{fig:BFluctuationsFFT_space}$c$. We can see the modes excited at $t\cdot s=0.5$ in $\delta B_z$ propagating quasi-parallel to $\langle \textbf{B}\rangle$, and most of the power concentrates around wavelengths $\sim 9$ $R_{L,i}^{\text{init}}$ in fig. \ref{fig:BFluctuationsFFT_space}$b$. Additionally, between $0.6\lesssim t\cdot s\lesssim 1.2$, the peak power oscillates around the parallel direction in a narrow cone, so the waves always propagate nearly parallel to $\langle \textbf{B}\rangle$, with most of the power concentrating at wavelengths $\sim 15$ $R_{L,i}^{\text{init}} \sim 21 R_{L,e}^{\text{init}}$ (for $m_i/m_e=2$), as shown in \textbf{fig. \ref{fig:BFluctuationsFFT_space}$c$}. 
The oblique character of firehose modes can also be seen in fig. \ref{fig:BFluctuationsFFT_space}$d$, where the spatial Fourier transform of $\delta B_z$ is shown at $t\cdot s=1.5$. In this case, most of the power is concentrated in modes with angles in the range $20^{\circ}\lesssim \theta_k \lesssim 65^{\circ}$.

\begin{figure}
    \centering
    \begin{tabular}{cc}
         \includegraphics[width=0.47\linewidth]{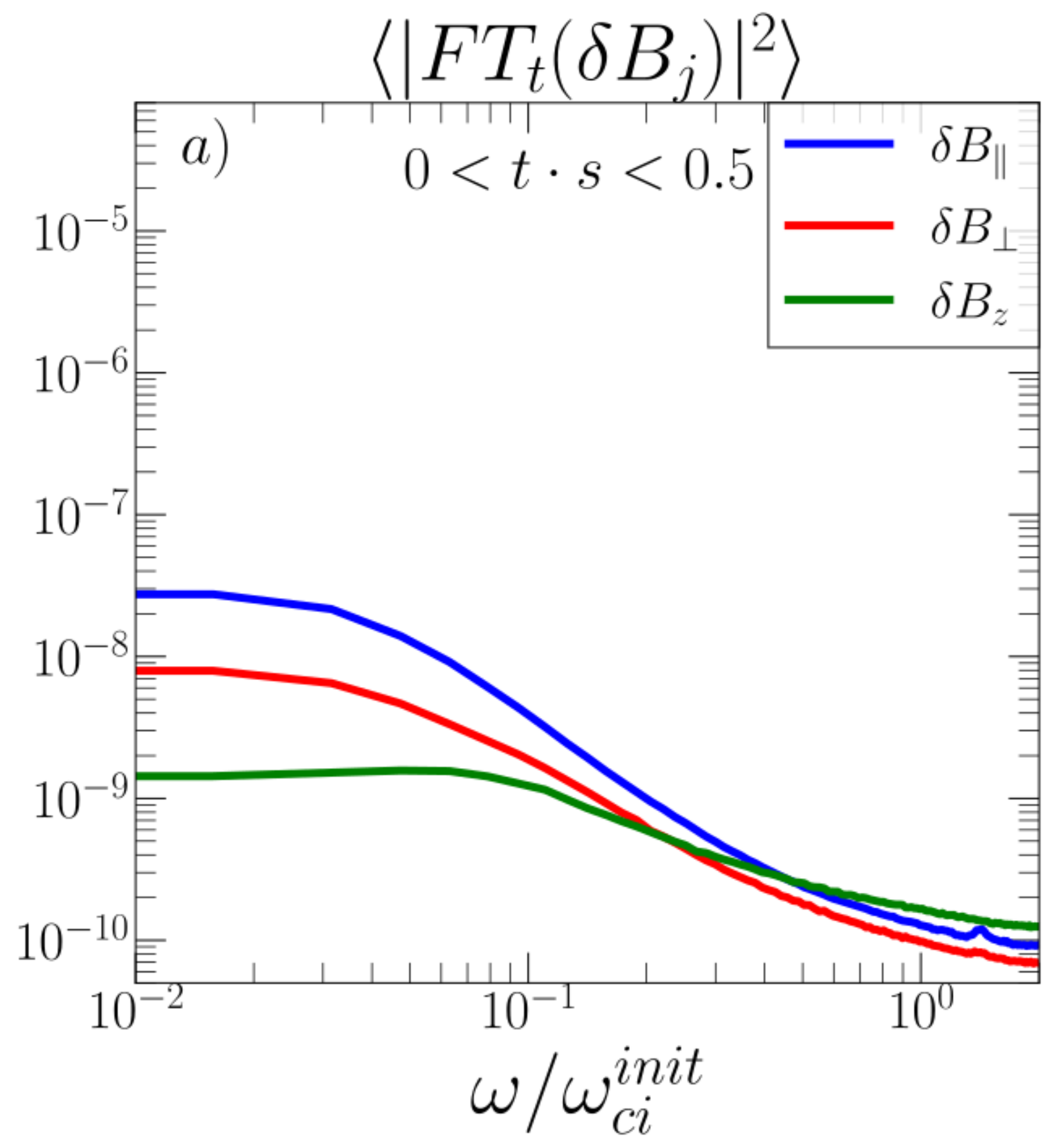}  &
         \hspace{-0.5cm}
         \includegraphics[width=0.47\linewidth]{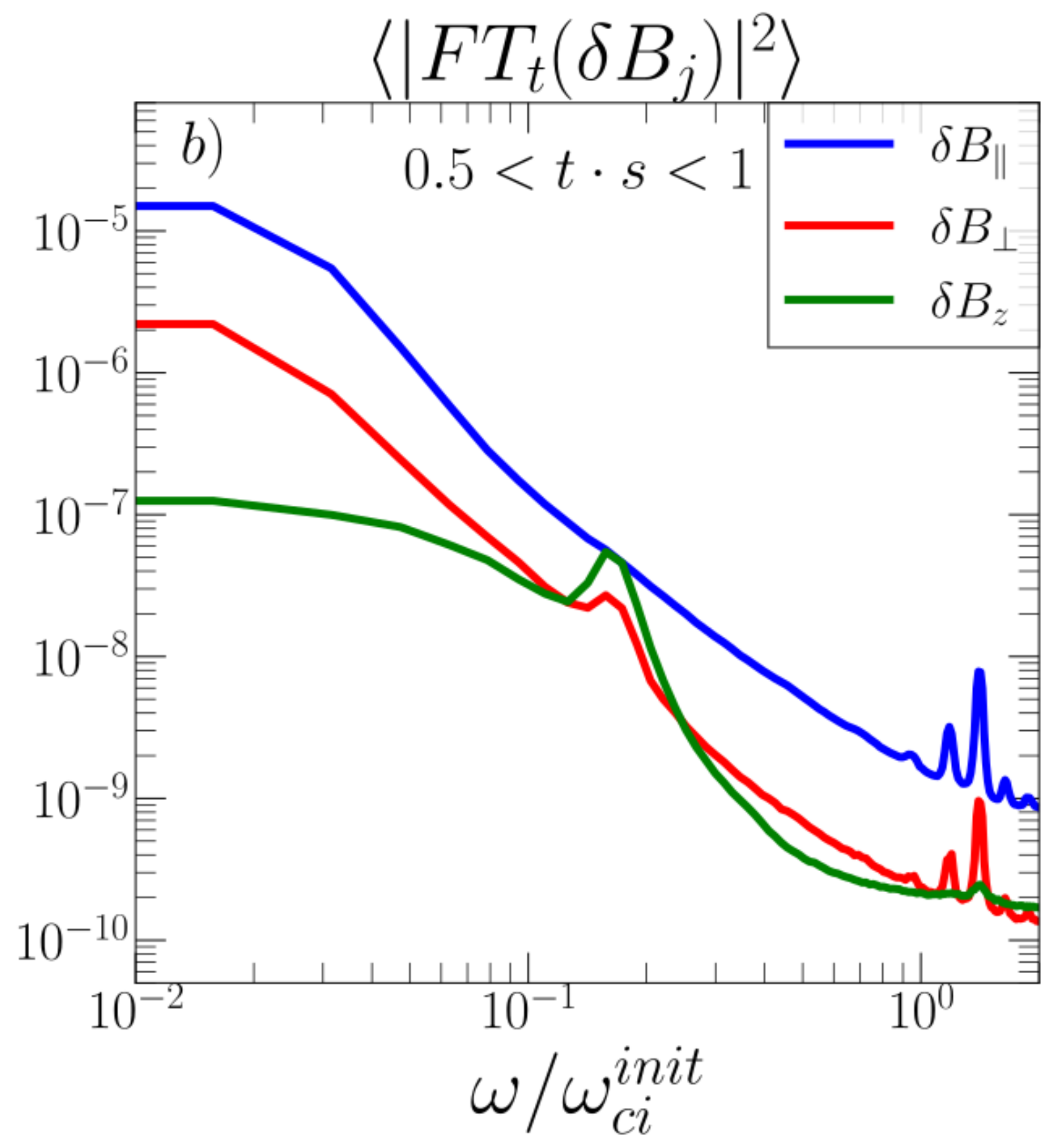}  \\
         \includegraphics[width=0.47\linewidth]{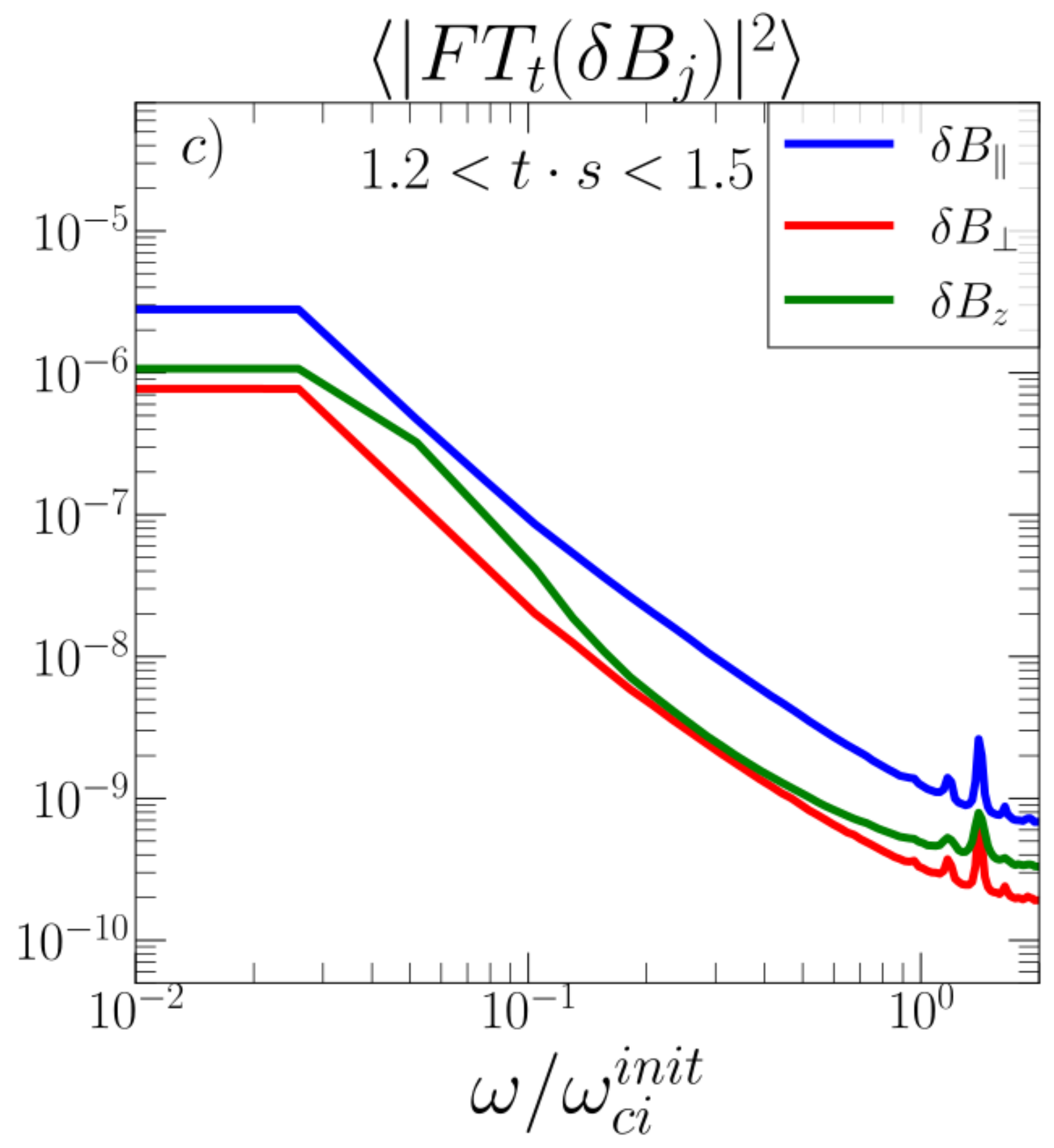} &
         \hspace{-0.5cm}
         \includegraphics[width=0.47\linewidth]{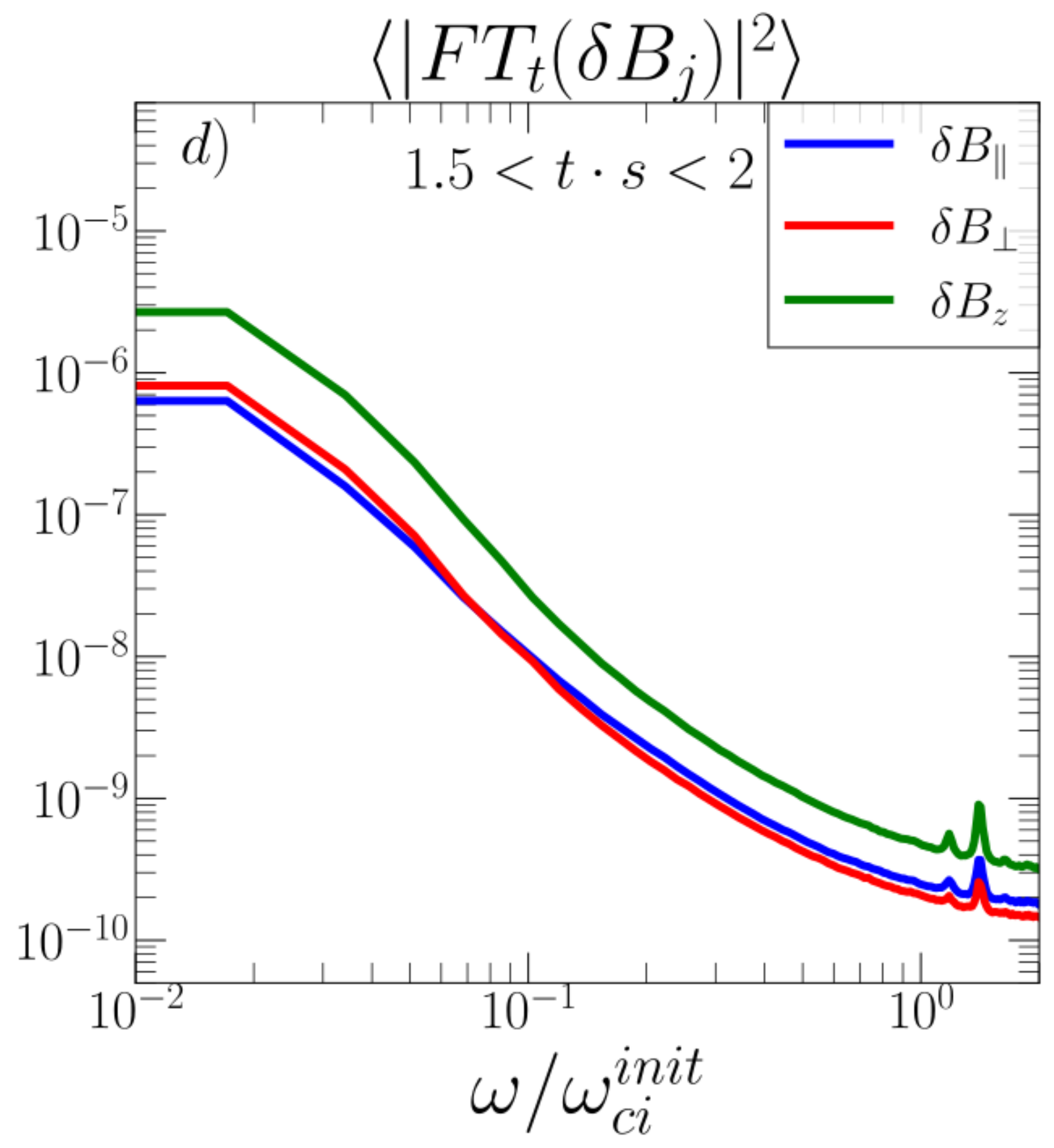} \\
    \end{tabular}
    \caption{Volume-averaged power spectrum in time of $\delta \textbf{B}$ along its three components, $\delta B_z$ (green line), $\delta B_{\perp}$ (red line) and $\delta B_{\parallel}$ (blue line) as a function of angular frequency for run Zb20m2w800 in a time interval $0 < t\cdot s < 0.5$ (panel $a$), $0.5 < t\cdot s < 1$ (panel $b$), $1.2 < t\cdot s < 1.5$ (panel $c$) and $1.5 < t\cdot s < 2$ (panel $d$). The frequencies are normalized by the initial cyclotron frequency $\omega_{c,i}^{\text{init}}$ of the ions.}
    \label{fig:BFluctuationsFFT_time}
\end{figure}

Finally, in fig. \ref{fig:BFluctuationsFFT_time} we show the Fourier transform in time averaged over the simulation box of the three components of $\delta \textbf{B}$, namely $FT_t(\delta B_{\parallel})$, $FT_t(\delta B_{\perp})$ and $FT_t(\delta B_z)$ for run Zb20m2w800 in four time intervals. During $0<t\cdot s<0.5$ (panel \ref{fig:BFluctuationsFFT_time}$a$), we can see that most of the power is concentrated at low frequencies, especially in $\delta B_{\parallel}$, consistent with the nature of mirror modes (\cite{Pokhotelov2002,Pokhotelov2004}). During $0.5<t\cdot s<1$ (panel \ref{fig:BFluctuationsFFT_time}$b$), the power at low frequency still dominates, consistent with the dominance of mirror modes in this interval, but also the transverse components $\delta B_{\perp,xy}$ and $\delta B_z$ exhibit a subdominant, broad resonant feature at $\omega \sim 0.2 \omega_{c,i}^{\text{init}} \sim 0.1\omega_{c,e}^{\text{init}}$ (for $m_i/m_e=2$), whereas $\delta B_{\parallel}$ and $\delta B_{\perp,xy}$ exhibit narrower resonances around $\omega \sim \omega_{c,i}^{\text{init}}$. The broad resonant feature at $\omega \sim 0.1\omega_{c,e}^{\text{init}}$ in both perpendicular components of $\delta \textbf{B}$ is consistent with the expected frequency of whistler lion roars (\cite{Baumjohann1999}). The nature \textbf{of} the narrower resonances at $\omega \sim \omega_{c,i}^{\text{init}}$ in $\delta B_{\parallel}$ is less clear, but could be related to the resonant nature of mirror instability via Landau resonances. During $1.2<t\cdot s<1.5$ (panel \ref{fig:BFluctuationsFFT_time}$c$), we can see that the broad resonance at $\omega \sim 0.1\omega_{c,e}^{\text{init}}$ is not present anymore, but the narrow resonant peaks still appear, although with less power. During $1.5<t\cdot s<2$ (panel \ref{fig:BFluctuationsFFT_time}$d$), we can see that the power in $\delta B_z$ dominates in power and resides at low frequencies, consistent with the nature of oblique firehose modes (\cite{Hellinger2000}). There are also narrow peaks at $\omega \sim \omega_{c,i}^{\text{init}}$, which is consistent with oblique firehose instability being resonant via cyclotron resonances (\cite{Hellinger2008}).


\subsection{Mass Ratio and Magnetization Dependence}
\label{sec:Anisotropy_mime_wcis_Comparison}
In this section, we explore the dependency of the heating on the mass ratio of ions to electrons, $m_i/m_e$ and the magnetization parameter $\omega_{c,i}^{\text{init}}/s$. 

\begin{figure}[hbtp]
    \centering
    \begin{tabular}{c}
         \includegraphics[width=\linewidth]{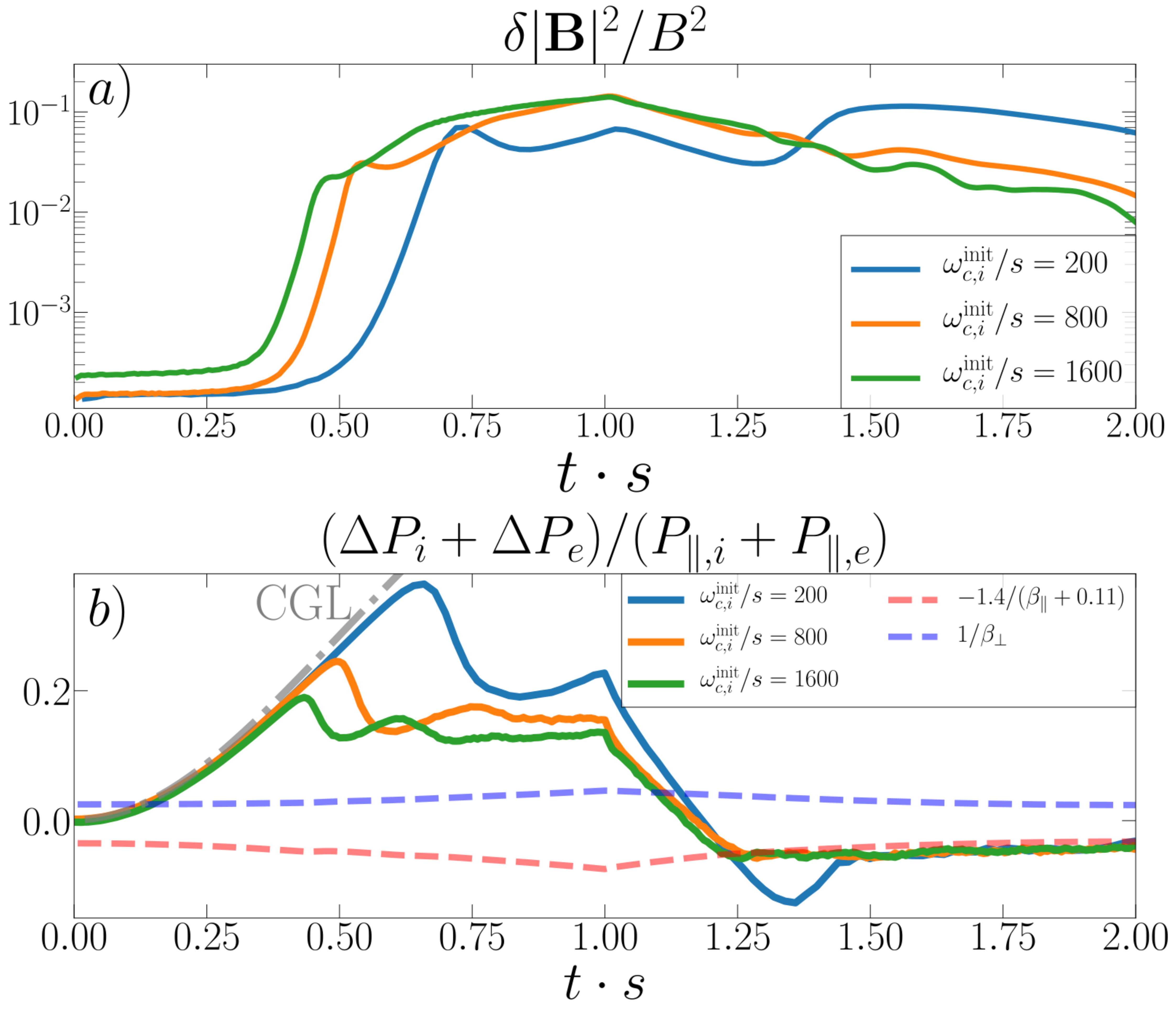} \\
         \includegraphics[width=\linewidth]{MagnetizationCompareHeating_rgb_ts2.pdf} \\
         \includegraphics[width=\linewidth]{Magnetization_lecs_CompareHeating_rgb_ts2.pdf}
    \end{tabular}
    \caption{Comparison between different magnetizations $\omega_{c,i}^{\text{init}}/s$. Panel $(a)$: the evolution of the magnetic field fluctuations $\delta |\textbf{B}|^2$ for runs Zb20m2w200, Zb20m2w800 and Zb20m2w1600, all with same mass ratio $m_i/m_e=2$ and magnetizations of $\omega_{c,i}^{\text{init}}/s=200$ (blue line), $\omega_{c,i}^{\text{init}}/s=800$ (orange line) and $\omega_{c,i}^{\text{init}}/s=1600$ (green line). Panel $(b):$ Evolution of the total pressure anisotropy $(\Delta P_i+\Delta P_e)/(P_{\parallel,i}+P_{\parallel,e})$ for the same runs as in panel $(a)$. The dashed lines show the approximate linear thresholds for the excitation of mirror (dashed blue line) and firehose (dashed red line) instabilities, and the dashed-dotted gray line shows the expected double-adiabatic evolution of the pressure anisotropy. Panel $(c):$ \textbf{(same as fig. \ref{fig:HeatingDependency}$c$)} the integrated ion gyroviscous heating for the same runs as in panel $(a)$ is shown in solid lines (cf. eq. \ref{eq:GyroviscousHeating}) for magnetizations $\omega_{c,i}^{\text{int}}/s=200$ (blue line), $\omega_{c,i}^{\text{int}}/s=800$ (red line) and $\omega_{c,i}^{\text{int}}/s=1600$ (green line). The dashed lines show the ion energy gain $\Delta U_i$ after being corrected for numerical heating. Panel $(d):$ (same as fig. \ref{fig:HeatingDependency}$d$) same quantities as in panel (c) but for electrons.}
    \label{fig:MagnetizationComparison}
\end{figure}

Figure \ref{fig:MagnetizationComparison}$a$ shows the evolution of the total magnetic fluctuations $\delta |\textbf{B}|(t)^2$ in one cycle $(0<t \cdot s<2)$ for simulations with $\omega_{c,i}^{\text{init}}/s=200, 800$ and $1600$ (runs Zb20m2w200, Zb20m2w800 and Zb20m2w1600 in Table \ref{table:SimulationParameters}). In all three simulations the particles have the same physical parameters other than the magnetization: $m_i/m_e=2, k_BT_i = k_BT_e = 0.1m_ic^2, \beta_i^{\text{init}}=20$. We can see that overall the evolution shows the same behavior; exciting mirror modes in the first half-cycle and firehose modes in the second half-cycle, and mirror modes are excited at relatively earlier times for increasing magnetizations\footnote{Note that, in our simulations, the larger the magnetization, the smaller the numerical value of the shear frequency, so the shear occurs more slowly with respect to the initial ion cyclotron frequency, therefore providing more time for the instabilities to develop. This translates to the apparent earlier excitation of the instabilities for larger magnetizations in fig. \ref{fig:MagnetizationComparison}$a$.}. Mirror fluctuations saturate around the same value for all runs, whereas for firehose modes it decreases with increasing magnetizations.

Figure \ref{fig:MagnetizationComparison}$b$ shows the pressure anisotropy evolution for the same set of simulations as in fig. \ref{fig:MagnetizationComparison}$a$. The overall evolution is similar in all cases; both mirror and firehose instabilities are excited and have time to limit the anisotropy growth. The main difference is reflected in the value that the anisotropy can reach before starting to get regulated by the instabilities. We will refer to this point as the overshoot in $\Delta P$. This can be seen more clearly at the beginning of each cycle (e.g. $t\cdot s \approx 0.5$), where the overshoot that $\Delta P/P_{\parallel}$ undergoes is largest for $\omega_{c,i}/s = 200$ and decreases for larger values of magnetization. This also happens when the shear motion is reversed, right before the firehose modes act and regulate the pressure anisotropy (e.g. $t \cdot s \approx 1.3$). This effect has been shown and discussed in previous works for both hybrid and fully kinetic PIC simulations in the context of a continuous shear motion (\cite{Kunz2014}, \cite{Riquelme2015}, \cite{Riquelme2018}). This difference in the overshoot peak value implies that, for lower magnetizations, mirror and firehose modes are less efficient in pitch-angle scattering the particles, taking longer to grow (in units of $s^{-1}$), and this allows $\Delta P/P_{\parallel}$ to keep growing to larger values until it effectively saturates due to scattering (\cite{Riquelme2015}). 

For completeness, in figs. \ref{fig:MagnetizationComparison}$c$ and \ref{fig:MagnetizationComparison}$d$ we show the same plots as in figs. \ref{fig:HeatingDependency}$c$ and \ref{fig:HeatingDependency}$d$, respectively, i.e. the ion and electron energy gain $\Delta U_i, \Delta U_e$ (dashed lines) after being corrected by numerical heating (see appendix \ref{sec:NumericalHeating}) and the integrated gyroviscous heating rate as a function of time (solid lines, cf. eq. \ref{eq:GyroviscousHeating}) for the same runs as in fig. \ref{fig:MagnetizationComparison}$a$. We can see that in all cases a net heating is obtained in each pump cycle, with the difference that the heating rate decreases for larger magnetizations. The reason for the decrease in the heating rate for larger magnetizations is directly related to the decrease of the peak value of $\Delta P/P_{\parallel}$ during the overshoot and also to the anisotropy decrease in the secular mirror modes regime. Indeed, the heating provided by the pressure anisotropy during this phase is retained after the pump cycle and, looking at eq. \ref{eq:GyroviscousHeating}, it will be larger for larger values of the anisotropy. This sets a direct dependence of the efficiency of gyroviscous heating on the overshoot and secular values that $\Delta P/P_{\parallel}$ produces during a pump cycle for different values of $\omega_{c,i}/s$. This way, higher shear frequencies (i.e. smaller $\omega_{c,i}/s$) will produce more heating, whereas low shear frequencies (large $\omega_{c,i}/s$) will produce less heating. In a turbulent medium like the ICM, where the plasma can be driven by a spectrum of turbulent eddies with different frequencies, the contribution of all these frequencies can be relevant in the total amount of heating the plasma can get via magnetic pumping. 

\begin{figure}[htbp]
    \centering
    \begin{tabular}{c}
         \includegraphics[width=\linewidth]{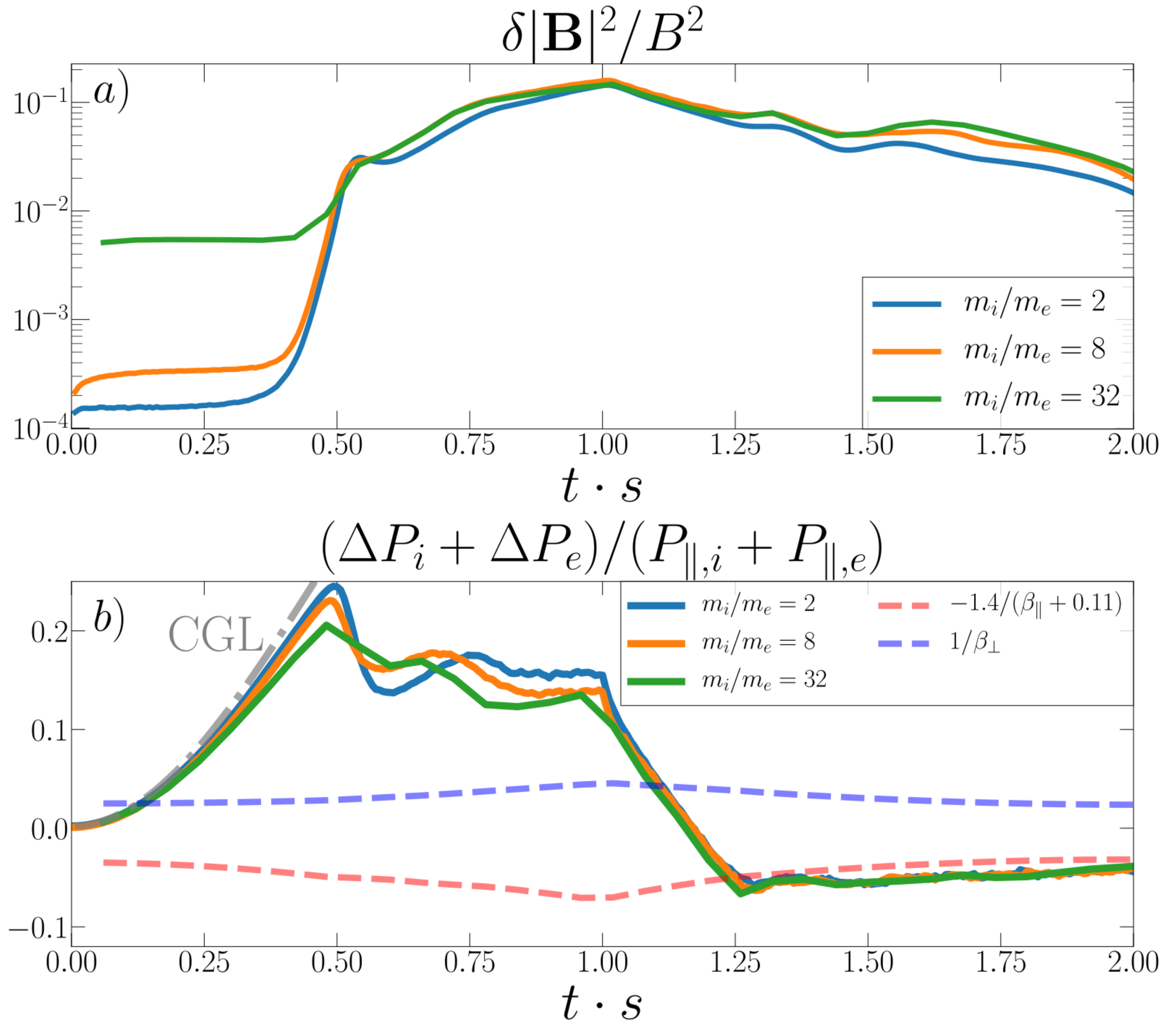} \\
         \includegraphics[width=\linewidth]{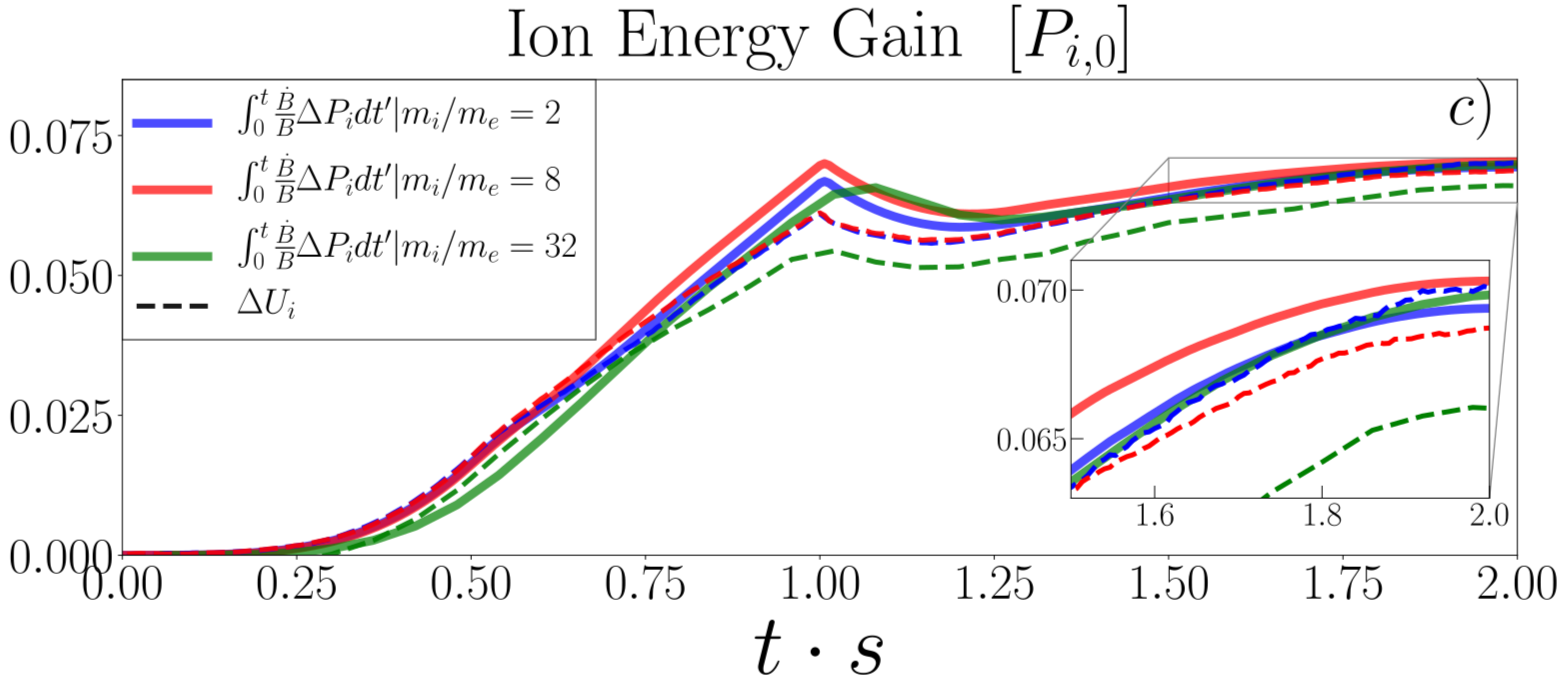} \\
         \includegraphics[width=\linewidth]{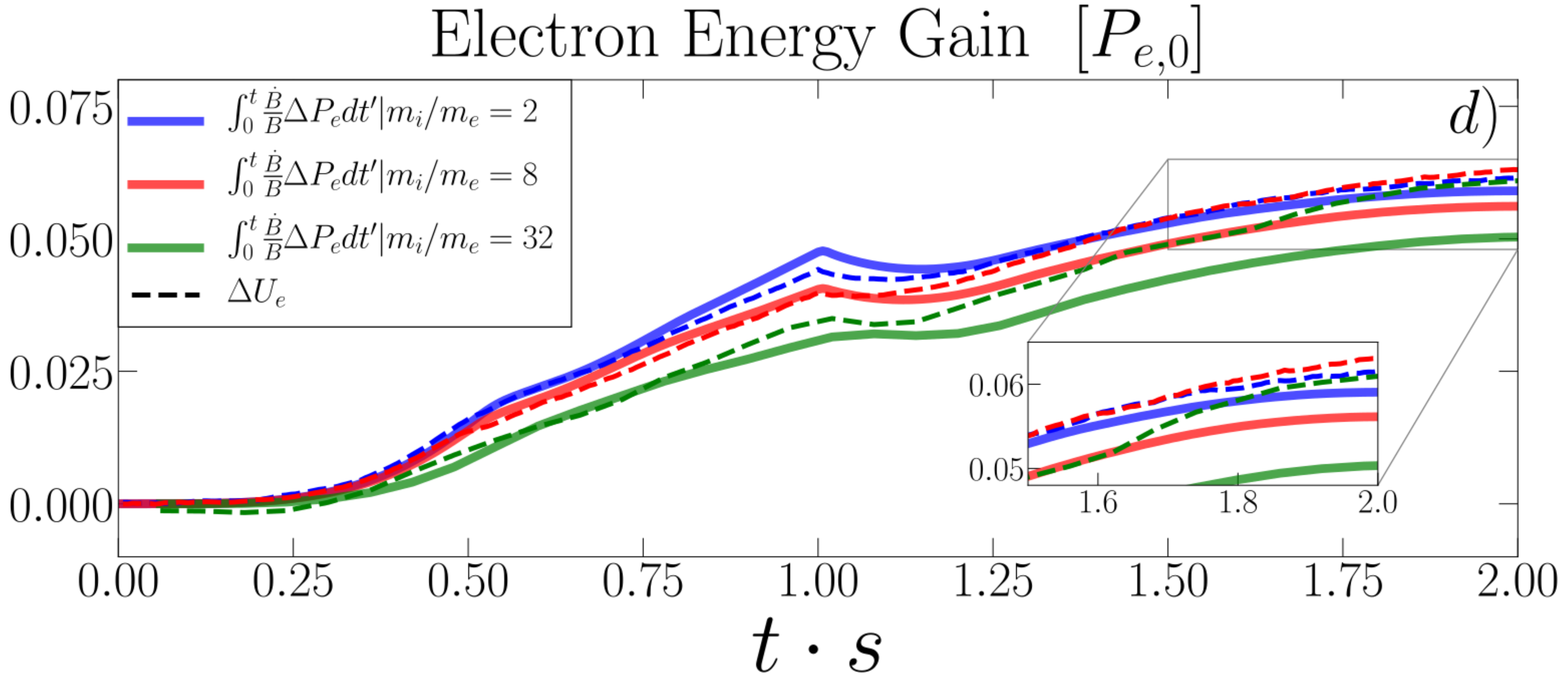}
    \end{tabular}
    \caption{Comparison between different ion-to-electron mass ratio $m_i/m_e$. Panel $(a)$: the evolution of the magnetic field fluctuations $\delta |\textbf{B}|^2$ for runs Zb20m2w800, Zb20m8w800 and Zb20m32w800, all with same magnetization $\omega_{c,i}/s=800$ and mass ratio of $m_i/m_e=2$ (blue line), $m_i/m_e=8$ (orange line) and $m_i/m_e=32$ (green line). Panel $(b):$ Evolution of the total pressure anisotropy $(\Delta P_i+\Delta P_e)/(P_{\parallel,i}+P_{\parallel,e})$ for the same runs as in panel $(a)$. The dashed lines show the approximate linear thresholds for the excitation of mirror (dashed blue line) and firehose (dashed red line) instabilities, and the dashed-dotted gray line shows the expected double-adiabatic evolution of the pressure anisotropy. Panel $(c):$ \textbf{(same as fig. \ref{fig:HeatingDependency}$a$)} the integrated ion gyroviscous heating for the same runs as in panel $(a)$ is shown in solid lines (cf. eq. \ref{eq:GyroviscousHeating}). The dashed lines show the ion energy gain $\Delta U_i$ after being corrected by numerical heating. Panel $(d)$: \textbf{(same as fig. \ref{fig:HeatingDependency}$b$)} same quantities as in panel $(c)$ but for electrons.}
    \label{fig:MassRatioComparison}
\end{figure}

Complementary to fig. \ref{fig:MagnetizationComparison}, fig. \ref{fig:MassRatioComparison}$a$ shows the evolution of the total magnetic fluctuations $\delta |\textbf{B}|(t)^2$ in one cycle $(0<t\cdot s<2)$ now for simulations with $m_i/m_e=2,8$ and $32$ (runs Zb20m2w800, Zb20m8w800 and Zb20m32w800 in Table \ref{table:SimulationParameters}). The three simulations share the same physical parameters: $\omega_{c,i}^{\text{init}}=800, k_BT_i = k_BT_e = 0.1m_ic^2, \beta_{i}^{\text{init}}=20$. We can see a similar behavior in all three cases; the mirror modes saturate at the same level in the first half of the pump cycle, whereas the firehose modes saturates at a slightly higher level for increasing mass ratio.

Figure \ref{fig:MassRatioComparison}$b$ shows the pressure anisotropy evolution for the same set of simulations as in fig. \ref{fig:MassRatioComparison}$a$. In this case, the evolution is very similar for all three simulations; in the first half of the cycle ($0<t\cdot s<1$) and after a brief double-adiabatic (CGL) evolution, they reach a similar overshoot level (as they all have the same magnetization) and start to be regulated by the interaction with mirror modes. In the second half of the cycle ($1<t\cdot s<2$) the anisotropies start to increase in absolute value and quickly start to be regulated by firehose modes, after reaching an overshoot that is also very similar in all three mass ratio cases. 

For completeness, in figures \ref{fig:MassRatioComparison}$c$ and \ref{fig:MassRatioComparison}$d$ we show the same plots as in figures \ref{fig:HeatingDependency}$a$ and \ref{fig:HeatingDependency}$b$. As discussed in section \ref{sec:Results}, we can see that in all three cases we obtain a net heating after one pump cycle, and its efficiency is very similar in all three simulations. This is consistent with the evolution of the pressure anisotropy as seen in fig. \ref{fig:MassRatioComparison}$b$ in terms of the overshoot level in the first and second halves of the pump cycle. The same happens for electrons; they are also heated after one pump cycle, and the net heating at the end of one cycle is slightly lower for larger mass ratio. In the case of electrons, however, given the limited range in mass ratio, we do not fully capture the electron-scale instabilities that could be present in a realistic scenario and that could also participate in the pitch-angle scattering of electrons.

Thus, this analysis shows that the action of gyroviscous heating on the ions is fairly independent of $m_i/m_e$, and its efficiency depends on $\omega_{c,i}^{\text{init}}/s$, where for larger magnetizations the retained heating becomes smaller.

\begin{figure}[hbtp]
    \centering
    \includegraphics[width=\linewidth]{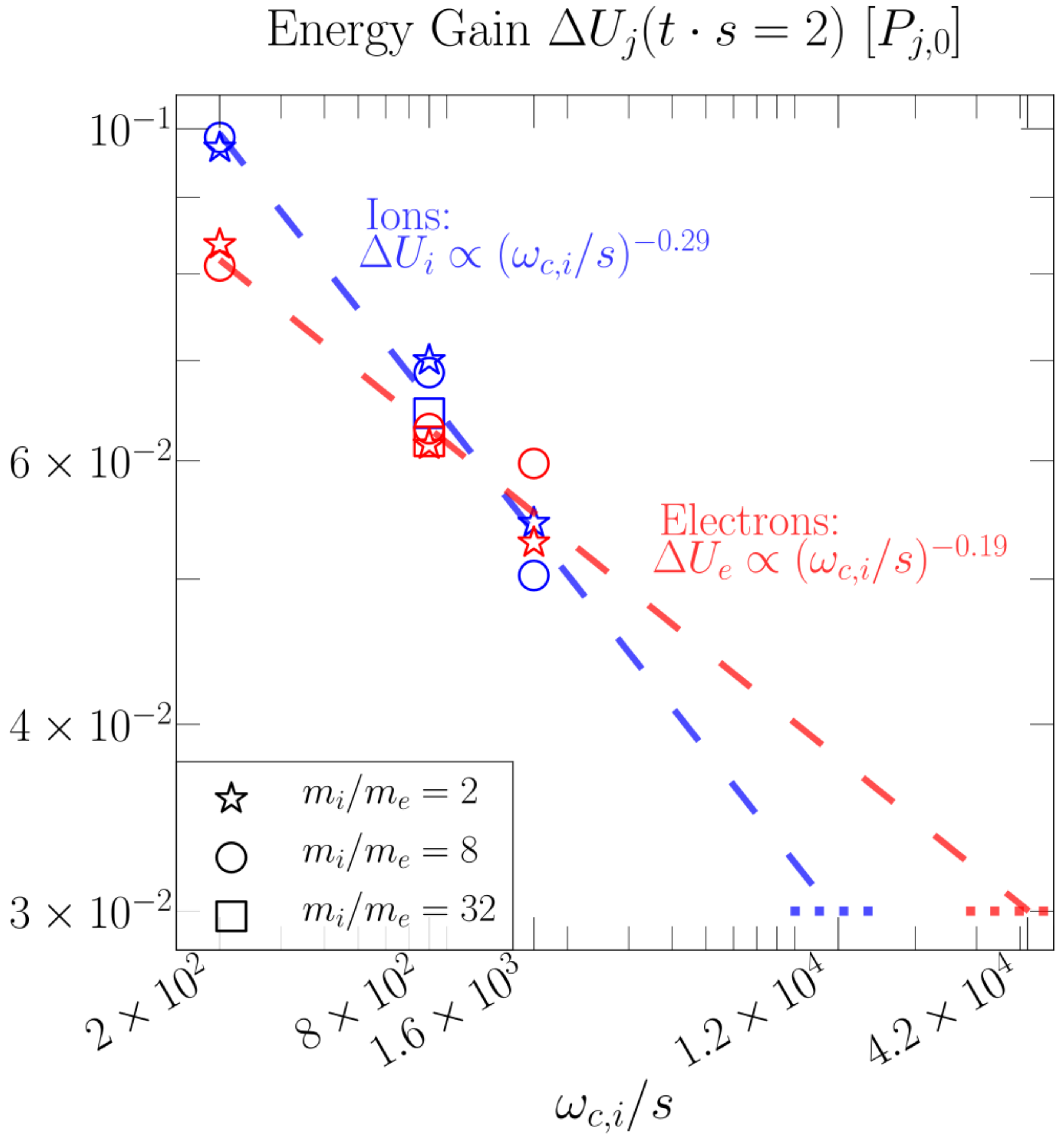}
    \caption{Particles' energy gain after 1 pump cycle ($t\cdot s$=2) as a function of magnetization for different mass ratios. The ion's and electron's energy gain are shown in blue and red symbols, respectively. Mass ratios of $m_i/m_e=2,8$ and $32$ are shown in open stars, circles and squares, respectively. The ion energy gain scales as $\Delta U_i \propto (\omega_{c,i}^{\text{init}}/s)^{\alpha_i}$ with $\alpha_i = -0.29 \pm 0.02$ (dashed blue line) whereas for electrons $\Delta U_e \propto (\omega_{c,i}^{\text{init}}/s)^{\alpha_e}$ with $\alpha_e = -0.19 \pm 0.02$ (dashed red line). The energy gained assuming marginal stability according to eq. \ref{eq:k20113} is shown in horizontal dotted blue line for ions and horizontal dotted red line for electrons, at the expected magnetization value according to the scaling relations above. All quantities are normalized by the species' initial pressure $P_{j,0} \ (j=i,e)$.} 
    \label{fig:EnergyGainScaling}
\end{figure}

Figure \ref{fig:EnergyGainScaling} summarizes our findings. The energy gain of ions (blue symbols) and electrons (red symbols) after 1 pump cycle are shown as a function of magnetization for different mass ratios. We obtain scaling relations $\Delta U_i \propto (\omega_{c,i}^{\text{init}}/s)^{\alpha_i}$, with $\alpha_i=-0.29\pm 0.02$ for ions and $\Delta U_e\propto(\omega_{c,i}^{\text{init}}/s)^{\alpha_e}$, with $\alpha_e = -0.19\pm 0.02$ for electrons. These relations can be used to estimate the heating rate of a turbulent spectrum of different frequencies (or shear rates $s$). It is worth noting that, at first glance, we could conclude that these scaling relations would imply a negligibly small energy gain for sufficiently high magnetizations. This, however, is not the case, as the energy gain is expected to be greater or equal than the heating provided assuming marginal stability. Indeed, according to eq. \ref{eq:k20113}, without overshoot the energy gain $\Delta U/P_0$ is of the order of $3\%$ per cycle assuming marginal stability. We plot this level in fig.\ref{fig:EnergyGainScaling} as horizontal dashed blue and red lines for ions and electrons, respectively, at the expected value of magnetization $\omega_{c,i}^{\text{init}}/s$ according to the scaling relations above\footnote{In a real setup, however, one could still expect some overshoot in the anisotropy at these values of magnetization.}. These levels provide a lower limit to the scaling relations we have derived.

\subsection{Plasma $\beta_i^{{init}}$ Dependence}
\label{subsec:beta}

We also explored how the heating by gyroviscosity acts for the case of a lower plasma beta, such as the solar wind, where it could coexist with other mechanisms of thermal heating and/or nonthermal acceleration mediated by microinstabilities, such as resonant wave-particle interaction with ion-cyclotron waves or whistlers (\cite{Riquelme2017,Ley2019,Cerri2021,Riquelme2022}). 

\begin{figure}[hbtp]
    \centering
    \begin{tabular}{c}
        \includegraphics[width=\linewidth]{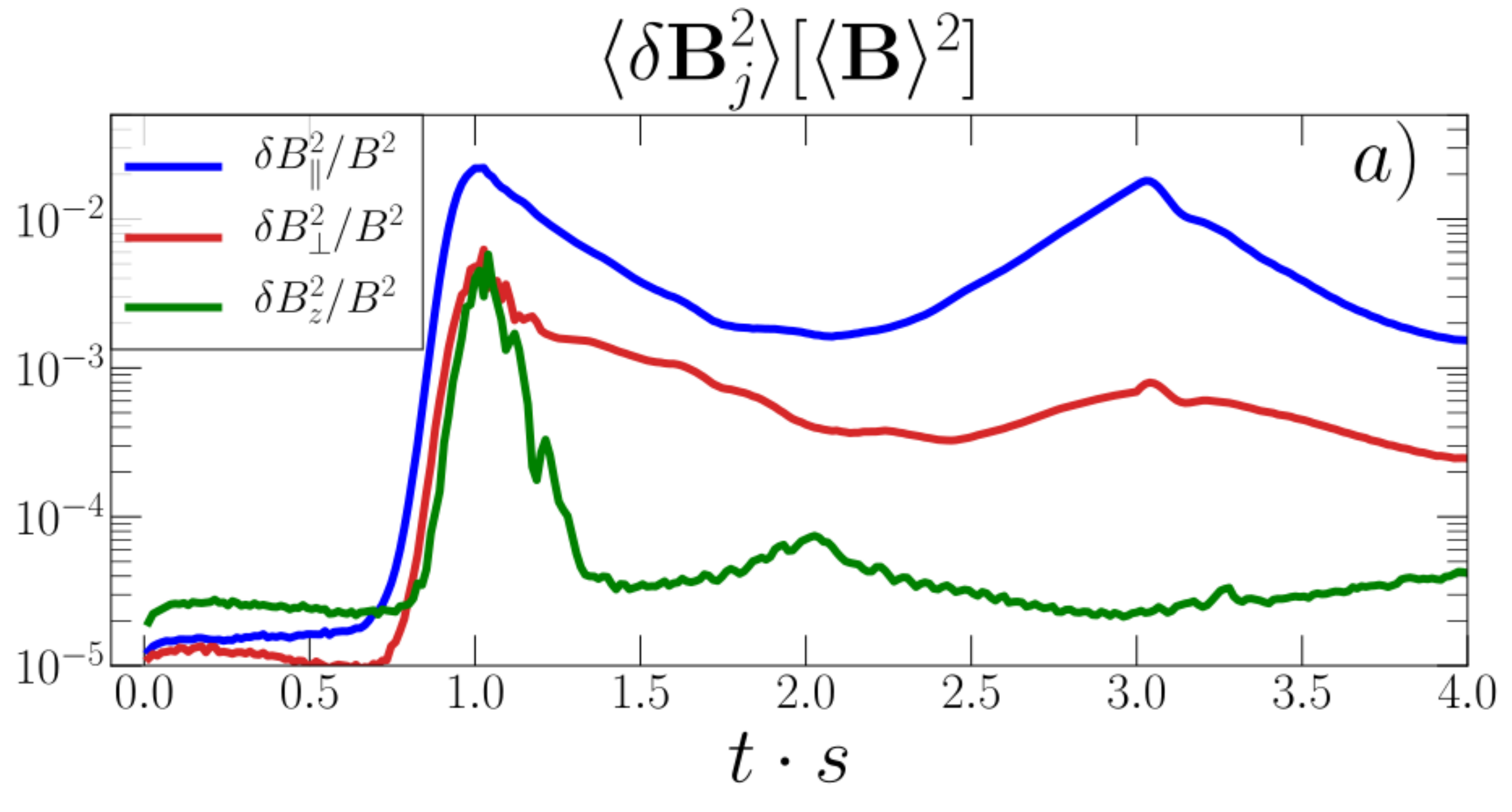} \\
        \includegraphics[width=\linewidth]{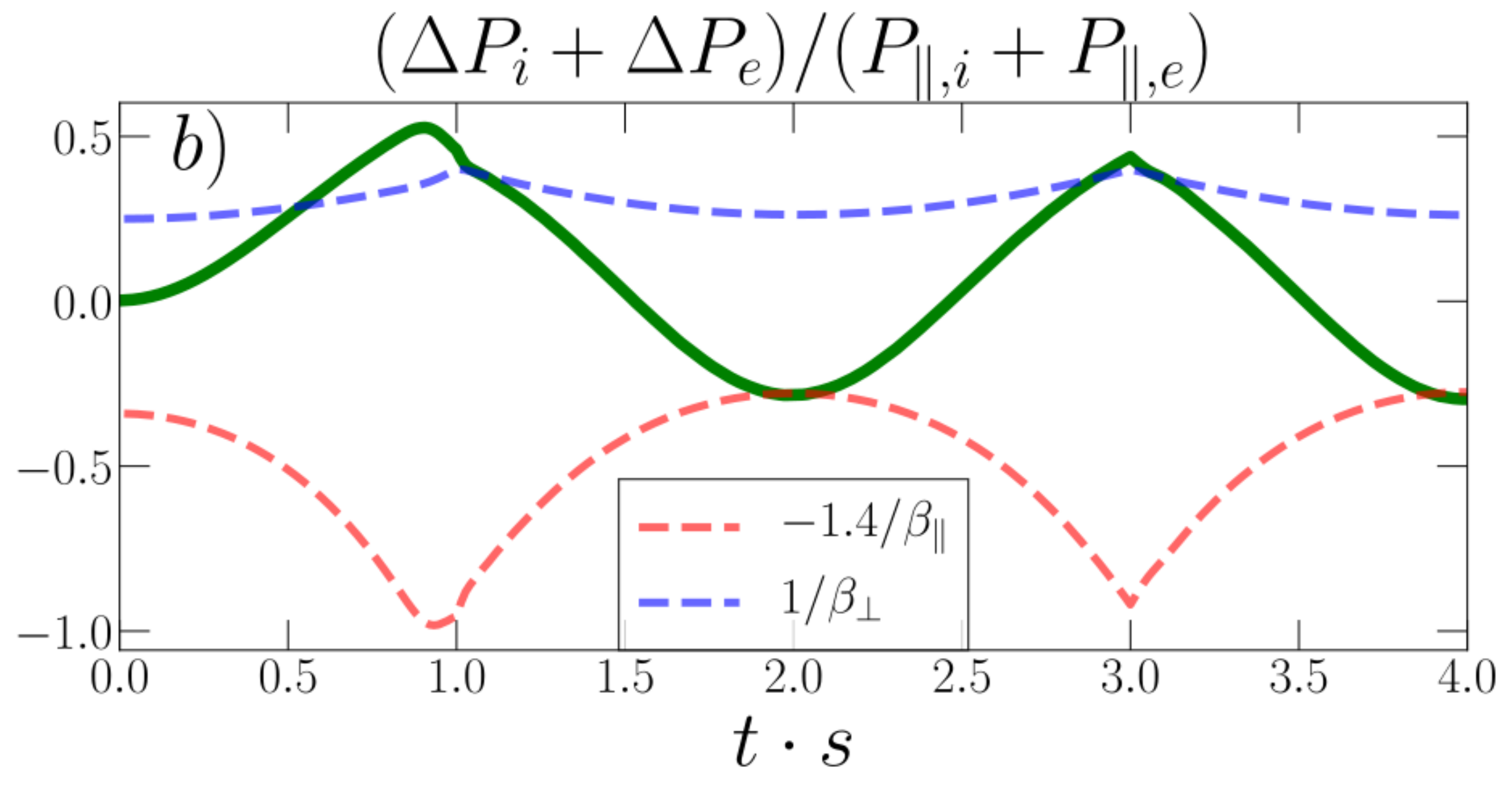} \\
        \includegraphics[width=\linewidth]{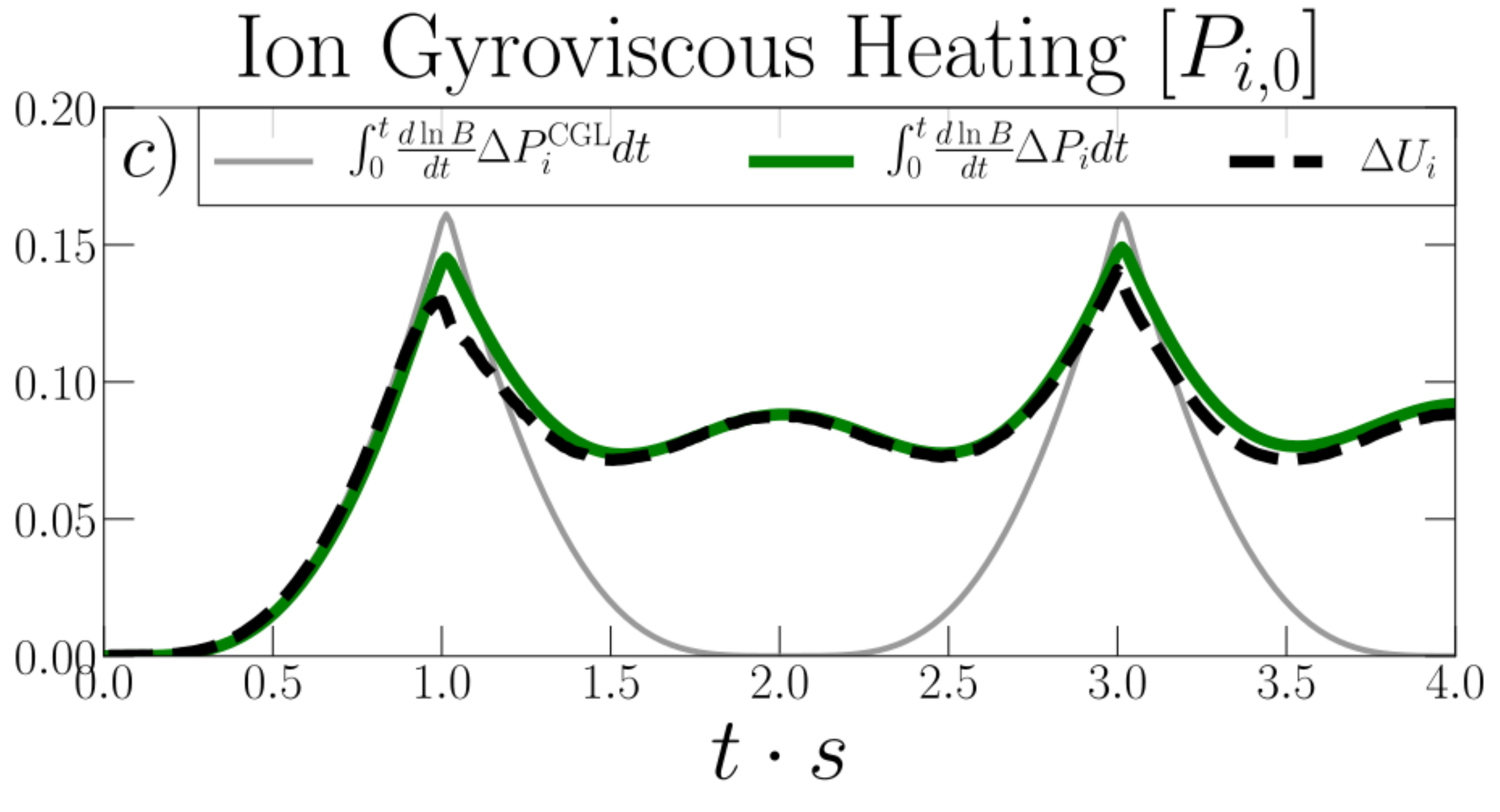} \\
    \end{tabular}
    \caption{Results of run Zb2m8w800 for initial $\beta_i^{\text{init}}=2$. Panel $(a)$: the three components of the volume-averaged magnetic energy in $\delta \textbf{B}$ as a function of time. Here $\delta B_{\parallel}$ (blue line) is the component parallel to $\langle \textbf{B} \rangle$, $\delta B_{\perp,xy}$ (red line) and $\delta B_z$ (green line) are, respectively, the components perpendicular to $\langle \textbf{B} \rangle$ in the plane and perpendicular to the plane of the simulation. Panel $(b)$: Evolution of the total pressure anisotropy $(\Delta P_i + \Delta P_e)/(P_{\parallel,i}+P_{\parallel,e})$. The dashed lines show the approximate linear thresholds for the excitation of mirror (dashed blue line) and firehose (dashed red line) instabilities, Panel $(c)$: the integrated ion gyroviscous heating is shown in solid green line and the ion energy gain $\Delta U_i$ after being corrected by numerical heating is shown in dashed black lines. The solid gray line shows the expected gyroviscous heating rate if the pressure anisotropy evolved according to CGL double-adiabatic prediction.} 
    \label{fig:beta2}
\end{figure}

Figure \ref{fig:beta2} shows the results of a periodic shear simulation with the same fixed physical parameters as the previous simulations except for the initial ion plasma beta: $\beta_i^{\text{init}}=2$ (run Zb2m8w800 in table \ref{table:SimulationParameters}). Figure \ref{fig:beta2}$a$ shows that, for a fixed value of magnetization ($\omega_{c,i}/s=800$), mirror modes take longer to be excited, reaching saturation around $t\cdot s \approx 1$ (in contrast to $t\cdot s\sim 0.4$ in the $\beta_i^{\text{init}}=20$ case), and consequently they do not have 
time to reach a nonlinear stage before the shear reverses. In the second half of the cycle ($1<t\cdot s<2$) we can see $\delta B_{\parallel}^2$ slowly decaying and $\delta B_z^2$ staying at very low amplitudes after a quick decay, exhibiting a slight rise towards $t\cdot s=2$. This means that firehose modes are not being excited sufficiently rapidly before the shear reverses, so they never gain sufficient energy to effectively scatter particles. This behavior is consistent with the $\beta^{-1}$ dependence of the pressure anisotropy instability thresholds of both instabilities, which are higher in this case. This is explicitly shown in fig. \ref{fig:beta2}$b$, where we can see that the approximate threshold for mirror instability is surpassed at a later time than for higher beta ($t\cdot s \approx 0.5$). Analogously, we can see that in the second half of the cycle ($1<t\cdot s<2$) the anisotropy can barely meet the approximate threshold for the excitation of firehose modes, consistent with the evolution of $\delta B_z^2$ in fig. \ref{fig:beta2}$a$.

The initial excitation of mirror modes in the first half of the pump cycle and the absence of firehose modes in the second half translates into an inefficient scattering process in one pump cycle, therefore we would not expect the evolution of the energy density to be very different from pure CGL evolution. Figure \ref{fig:beta2}$c$ shows the ion energy gain after being corrected by numerical heating and the integrated gyroviscous heating rate as a function of time. We see that, indeed, during the first half of the pump cycle ($0<t\cdot s<1$), mirror modes start to interact with ions and this somewhat extends to the second half of the cycle ($1<t\cdot s<2$). This allows retention of a small fraction of the energy injected by the shear motion. However, the absence of scattering provided by firehose modes does not allow the heating to continue growing, and the ion internal energy stays relatively constant. 

In the second cycle, the evolution differs from the previous one, and the cycles are no longer reproducible, unlike the high-beta case. The subsequent heating around $t\cdot s \approx 3$ is completely reversible as the scattering becomes inefficient, and there is no additional energy gain. We see that the efficiency of the gyroviscous heating by magnetic pumping is much lower for $\beta_i^{\text{init}}=2$. This is directly related to the inefficiency of the scattering processes during a pump cycle, and is set by the slower excitation of mirror and firehose instabilities with respect to the length of the cycle. 

It is important to note that for larger magnetization values, the instabilities would have more time to develop (relative to the shear frequency $s$) and to interact with the particles. The heating could therefore be relatively more efficient, although the anisotropy thresholds would be the same for any magnetization value for a given initial beta, setting a more stringent limiting factor for the overall efficiency. 

The results shown in this section set a lower, magnetization-dependent limit in $\beta$ for the action of this heating mechanism, therefore for moderate values of $\beta$ the heating rate is an increasing function of $\beta$. For large enough $\beta$, it is expected that the heating rate goes as $\beta^{-1}$ (\cite{Kunz2011}).

\section{Pitch Angle Scattering Model}
\label{sec:ScatteringModel}
A self-consistent treatment of the interaction of mirror and firehose modes with particles in a pump cycle would require solving equation \ref{eq:DriftKineticEquationScattering} coupled with equations for the scattering frequency $\nu$ assuming, for example, a quasilinear regime for each instability. This procedure turns out to be rather difficult, as the instabilities do not closely follow a quasilinear evolution (\cite{Hellinger2008,Hellinger2009,Rincon2014}), but reach nonlinear amplitudes in our simulations. Alternatively, we consider a simplified model similar to the bounded anisotropy model from \cite{Hellinger2008} where we directly evolve the scattering frequencies representing the interaction with mirror and firehose modes coupled to the evolution of the pressure anisotropy and plasma $\beta$. 

Assuming for simplicity one particle species and a nonrelativistic plasma, we can obtain the nonrelativistic evolution equation for $\Delta P$ by multiplying equation \ref{eq:DriftKineticEquationScattering} by $-pvP_2(\mu)$ and integrating over momentum space. In order to perform this integration, we expand the distribution function in Legendre polynomials $f(t,p,\mu)=\sum_n f(t,p)P_n(\mu)$, where $P_n(\mu)$ is the Legendre polynomial of order $n$. For our purposes here, it is sufficient to expand $f$ up to $n=2$, $f=f_0P_0+f_2P_2$, where $f_0$ is the isotropic part of the distribution function and $f_2$ naturally couples to $\Delta P$. This leads to:

\begin{align}
    \frac{d\Delta P}{dt} = \frac{3P-\Delta P}{B}\frac{dB}{dt} -3\nu\Delta P,
    \label{eq:Model_DeltaP}
\end{align}
where we assumed $\nu$ independent of $\mu$. We then characterize the scattering via waves excited by positive and negative pressure anisotropies, so $\nu = \nu_{+} + \nu_{-}$. Waves that are not excited are assumed to be damped and both damping and excitation processes are linear. This translates to the following pair of equations:

\begin{align}
    \frac{d\nu_+}{dt} &= \gamma_+ \nu_+ \left( \frac{\Delta P}{P} - \frac{\xi_+ }{\beta} \right) + \dot{\nu_b}, \label{eq:Model_nuplus}\\
    \frac{d\nu_-}{dt} &= -\gamma_- \nu_- \left( \frac{\Delta P}{P} - \frac{\xi_- }{\beta}\right) + \dot{\nu_b},  \label{eq:Model_numinus}
\end{align}
where $\beta=8\pi P/B^2$, $\gamma_+$,$\gamma_-$ are the wave growth rates, $\xi_+$,$\xi_-$ are the constants of order unity related to the linear instability thresholds introduced in eqn. \ref{eq:k20111} and $\dot{\nu_b}$ is a constant numerical parameter that we include to set the evolution on the right path without the final result being sensitive to its value, and is taken to be small.
Given the forms of equations \ref{eq:Model_nuplus} and \ref{eq:Model_numinus}, we can redefine equation \ref{eq:Model_DeltaP} to an equation for $A\equiv \Delta P/P$:

\begin{align}
    \frac{dA}{dt}=-3\nu A + \frac{3-A-\frac{2}{3}A^2}{B}\frac{dB}{dt}.
    \label{eq:Model_A}
\end{align}

Finally, using $U=3P/2$ along with equation \ref{eq:GyroviscousHeating} we obtain an evolution equation for $\beta$:

\begin{align}
    \frac{d\beta}{dt}=-2\beta\left( 1 - \frac{A}{3} \right)\frac{1}{B}\frac{dB}{dt}.
    \label{eq:Model_beta}
\end{align}

Equations \ref{eq:Model_nuplus}, \ref{eq:Model_numinus}, \ref{eq:Model_A} and \ref{eq:Model_beta} constitute the basic equations for the model. In order to compare with the results of our simulations, we prescribe the evolution of the magnetic field to be the same as in our previous PIC runs, and set $\gamma_{\pm}$ by fitting an exponential growth rate to the linear stages of mirror and firehose fluctuations; $\delta B_{\parallel}^2$ ($0.4<t\cdot s<0.52$) and $\delta B_z^2$ ($1.22<t\cdot s<1.31$), respectively (see fig. \ref{fig:DeltaP}). 
Importantly, the growth rates in our simulations need not be constant; the rates may depend on $\Delta P$ through the instability parameter $(\Delta P/P - \xi_{\pm}/\beta)$. To account for this dependence, we fit a function of the form:

\begin{align}
    \delta B^2_{\parallel, z} (t) \propto \exp(\Gamma_{\pm}t),
\end{align}
assuming constant $\Gamma_{\pm}$ over the fitted time interval.  Then, in our model equations \ref{eq:Model_nuplus}--\ref{eq:Model_numinus}, we take $\gamma_{\pm}$ equal to $\Gamma_{\pm}$ divided by the average of $(\Delta P/P - \xi/\beta)$ over the same fitted time interval and measured directly from the simulations.

Additionally, choosing to normalize $\nu$ by the initial cyclotron frequency, $\hat{\nu}=\nu/\omega_{c,i}^{\text{init}}$, $\gamma_{\pm}$ by the shear frequency, $\hat{\gamma}_{\pm}=\gamma_{\pm}/s$ and time by the inverse shear frequency, $\hat{t}=t/s^{-1}$, the equations remain unaltered except for the equation for $A$:

\begin{align}
    \frac{dA}{d\hat{t}}=-3\left(\frac{\omega_{c,i}^{\text{init}}}{s} \right) \hat{\nu} A + \frac{3-A-\frac{2}{3}A^2}{B}\frac{dB}{d\hat{t}}.
    \label{eq:Model_A_Norm}
\end{align}

It is worth noting that the magnetization parameter $\omega_{c,i}^{\text{init}}/s$ naturally appears in the first term of the RHS of equation \ref{eq:Model_A_Norm}, causing the same effect that we see in our simulations; a higher magnetization leads to a smaller overshoot and a faster regulation of the pressure anisotropy. 

\begin{figure}[hbtp]
    \centering
    \begin{tabular}{c}
         \includegraphics[width=\linewidth]{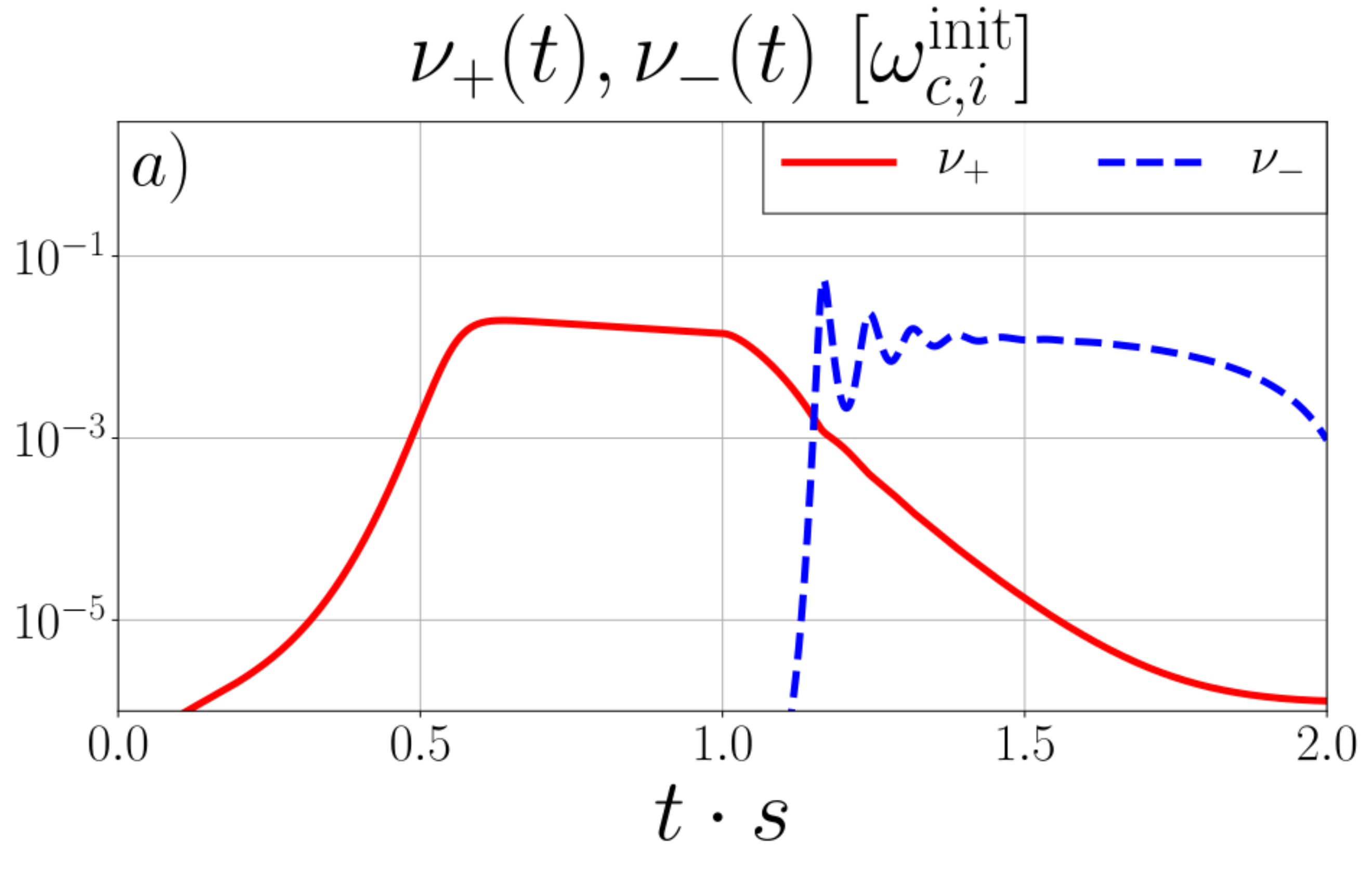}  \\
         \includegraphics[width=\linewidth]{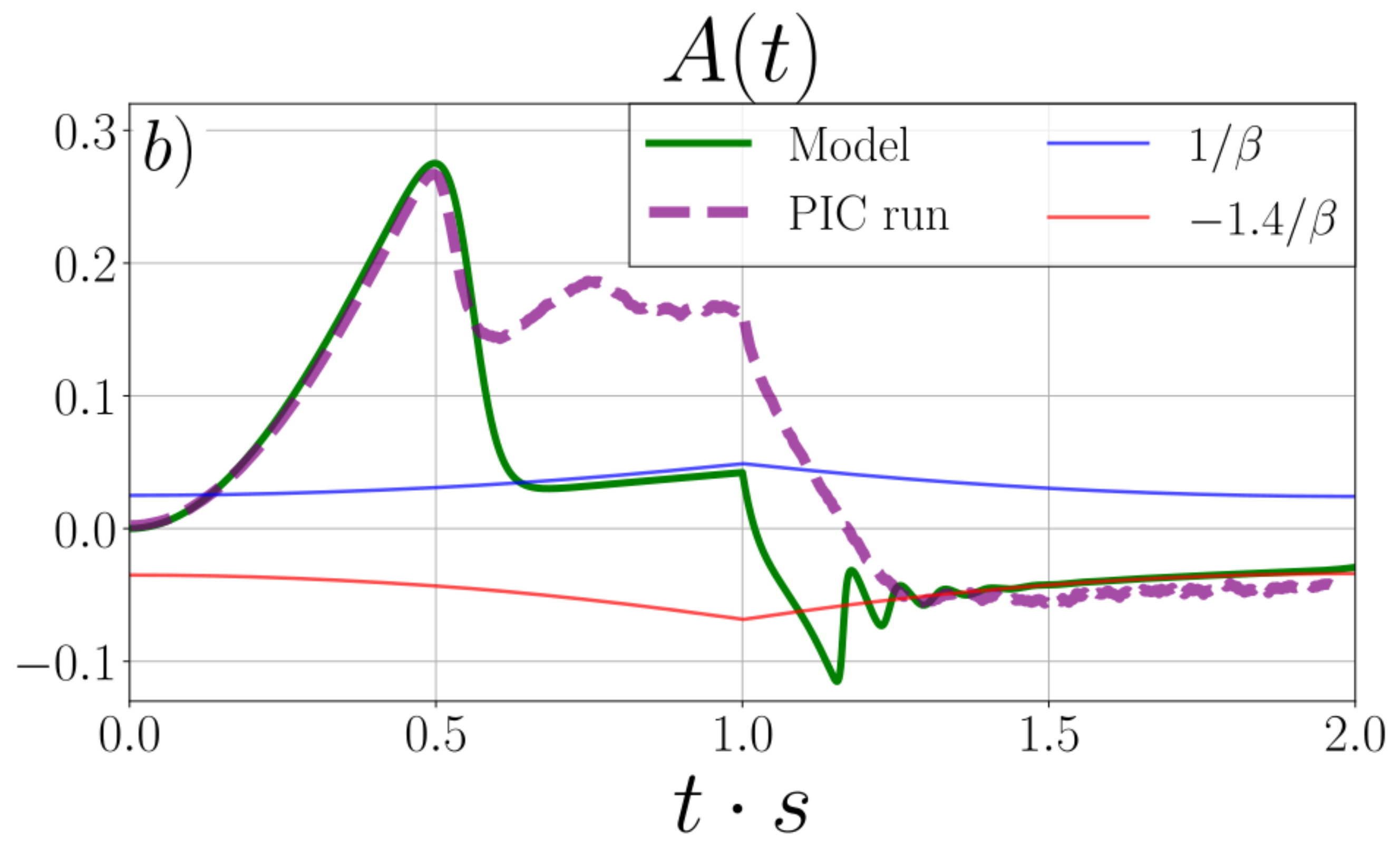}  \\
         \includegraphics[width=\linewidth]{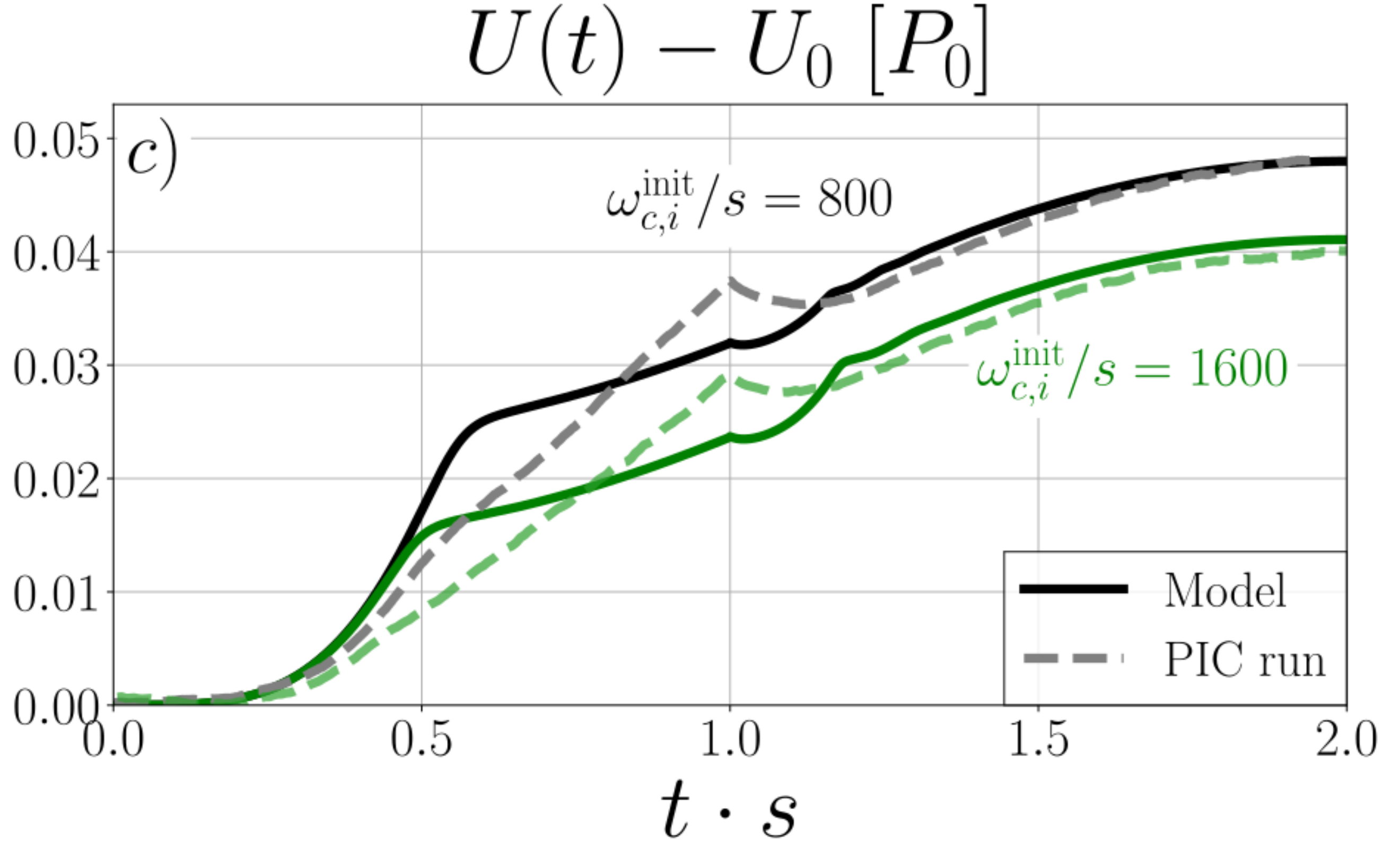}
    \end{tabular}
    \caption{Results of our pitch angle scattering model up to $t\cdot s = 2$. Panel $(a)$: The evolution of $\nu_+$ (solid red line) and $\nu_-$ (dashed blue line) normalized by the initial cyclotron frequency. Panel $(b)$: the evolution of $A(t)=\Delta P/P$ is shown in solid green line, and the dashed purple line shows the total anisotropy evolution $(\Delta P_i + \Delta P_e)/(P_i+P_e)$ for run Zb20m2w800. The solid blue line shows the mirror instability threshold $\xi_+/\beta$ whereas the solid red line shows the firehose instability threshold $-\xi_-/\beta$ (where we chose $\xi_+=1$ and $\xi_-=1.4$). Panel $(c)$: The energy gain $\Delta U=U-U_0$ calculated from U=3P/2 and $P=P_0 (B/B_0)^2(\beta/\beta_0)$ is shown in solid black line, and the dashed gray line shows the evolution of the total energy gain $\Delta U_i+\Delta U_e$ for run Zb20m2w800. Similarly, the solid green line shows the energy gain $\Delta U=U-U_0$ for a magnetization of $\omega_{c,i}^{\text{init}}/s=1600$, whereas the dashed green line shows the total energy gain $\Delta U_i+\Delta U_e$ for run Zb20m2w1600. Both quantities are normalized by the initial total pressure.}
    \label{fig:Model}
\end{figure}

In fig. \ref{fig:Model} we show the results for one pump cycle. We set $\omega_{c,i}^{\text{init}}/s=800$, $\gamma_+=152.18 s$, $\gamma_-=5406.17 s$, $\dot{\nu_b}=10^{-4} s$, $\xi_+=1$ (i.e. the approximate mirror linear threshold), $\xi_-=-1.4$ (i.e. the approximate oblique firehose threshold), and the initial conditions $\nu_{+,0}= \nu_{-,0}=0$, $A_0=0$, $P_0=0.4$ and $\beta_0=40$. In panels \ref{fig:Model}$b$ and \ref{fig:Model}$c$ we also include the respective results from our PIC simulation, run Zb20m2w800 (black lines). 

Additionally, in fig. \ref{fig:Model}$c$, we include the results for magnetization $\omega_{c,i}^{\text{init}}/s=1600$ (solid green line), which we compare with the PIC run Zb20m2w1600 (green dashed line). For this case, we set $\gamma_+= 218.96 s$, $\gamma_-= 1967.160s$, and the rest of the parameters and initial conditions are the same as in the $\omega_{c,i}^{\text{init}}/s=800$ case.

We can see that our model is phenomenologically consistent with our PIC simulations; the anisotropy is effectively regulated by the scattering frequencies $\nu_{\pm}$ that are alternately excited, as shown in fig. \ref{fig:Model}$a$: when the anisotropy $A(t)$ surpasses the given mirror threshold $\xi_+/\beta$ (solid green curve in fig. \ref{fig:Model}$b$), $\nu_+$ is excited, reaches a saturated value and subsequently decays when $B(t)$ start to decrease. Analogously, $\nu_-$ is not excited until the second half of the cycle, just when $A(t)$ surpasses the firehose threshold $-\xi_-/\beta$, reaching a similar saturated value as $\nu_+$ and regulating the anisotropy rapidly. It is interesting to note that, as $\nu_-$ is only excited when $A(t)$ surpasses the firehose threshold and the decrease of $B(t)$ continuously drives more anisotropy, this produces a series of $\nu_-$ excitations whenever $A(t)$ surpasses the firehose threshold (solid green curve in fig. \ref{fig:Model}$b$), generating an oscillatory behavior between stable and unstable regions. This is a distinctive feature of the oblique firehose instability related to its self-destructive properties (\cite{Hellinger2008,Hellinger2017}).

We can also see the presence of anisotropy overshoots in both mirror ($0<t\cdot s<1, A>0$) and firehose ($1<t\cdot s<2, A<0$) stages in fig. \ref{fig:Model}$b$ (solid green curve). The overshoot of $A(t)$ in the model during the mirror phase ($t\cdot s\approx 0.5$) coincides very well with our PIC run (dashed purple line), whereas in the firehose phase the overshoot is larger. Furthermore, we can see that $A(t)$ can be effectively pinned to the approximate mirror threshold $\xi_+/\beta$ in the model, in contrast to the nonlinear evolution of the anisotropy in PIC runs, which saturates at a larger value above the approximate mirror threshold. This feature is not captured by the model since it assumes linear wave excitation only, and demands further investigation on the nonlinear evolution of mirror modes and their conditions for marginal stability, in order to provide a more precise instability threshold. The discrepancy between the PIC simulations and the model may be evidence for a wave
physics effect, such as a nonlinear damping mechanism, such that a higher level of plasma anisotropy is required to maintain a steady state than predicted by linear theory.

The evolution of the energy gain $\Delta U$ is also consistent with our PIC runs, and heating is retained in each pump cycle, as evident in fig. \ref{fig:Model}$c$ (solid black curve). In the first half of the cycle, we can see that the energy gain increases following a CGL evolution until the anisotropy $A(t)$ is limited by the action of $\nu_+$ ($t\cdot s \approx 0.5$), after which it continues to increase but at a lower rate, given that $A(t)$ is now pinned to the mirror threshold $\xi_+/\beta$ at a lower amplitude. The energy gain in our PIC runs evolves similarly, but the change in slope is less pronounced, as the anisotropy saturates at a value not very far from the overshoot (see dashed yellow curve in fig. \ref{fig:Model}$b$). This also produces a slight difference in the energy gained by $t\cdot s=1$. In the second half of the cycle, the energy gain evolves very similarly to the PIC run; the difference both curves show at $t\cdot s=1$ is somewhat compensated by a larger gyroviscous cooling produced in the PIC run, given the larger amplitude at which the anisotropy saturates (see fig. \ref{fig:Model}$b$). Finally, by the end of the cycle, the energy gain from the model reaches a very similar value to the PIC run.


\section{Summary and Conclusions}
\label{sec:Conclusions}
In this work, we have studied how ions and electrons can be heated by gyroviscosity in a weakly collisional, high beta plasma subject to a magnetic pumping driving configuration where the scattering process is provided by mirror and oblique firehose fluctuations that grow from instabilities driven by pressure anisotropy $\Delta P\equiv P_{\perp}-P_{\parallel}$ in the plasma. By performing 2D PIC simulations of a periodically sheared domain we were able to self-consistently excite these instabilities during several pump cycles, allowing a detailed study of their linear and nonlinear stages and their respective roles in the heating mechanism. Our simulation parameters are given in Table \ref{table:SimulationParameters}. Because the plasma is not compressed, the heating is provided entirely by shear. However, gyroviscous heating has also been demonstrated in a compressing plasma at lower beta (\cite{Ley2019}), so we do not expect the restriction to pure shear to be a fundamental requirement. 

When initially $\beta_i^{\text{init}}=20$, the magnetization $\omega_{c,i}^{\text{init}}/s=800$, and $k_BT_i/m_ic^2=0.1$, we saw that the plasma can effectively retain about $40\%$ of the energy transferred during half a shear cycle in a CGL evolution ($t\cdot s=1$, solid gray curve in fig. \ref{fig:gyroviscousheating}). This corresponds to 2$-$3 times the rate that holds for marginally stable anisotropy with no overshoot. 
The efficiency of the heating is not strongly dependent on $m_i/m_e$ but does show a dependence on magnetization $\omega_{c,i}^{\text{init}}/s$; larger magnetization provides less heating per shear cycle. Physically, the dependence of the heating efficiency on magnetization is directly related to the evolution of the pressure anisotropy $\Delta P$ throughout the cycle, and consequently on the evolution of mirror and firehose instabilities. 
At the lowest magnetization we studied ($\omega_{c,i}^{\text{init}}/s=200$), $\Delta P/P$ overshoots the mirror instability threshold by more than 50\%, but the overshoot decreases with increasing magnetization. It is expected, however, that for large magnetizations, even with a very small overshoot, the heating reaches a non-zero value given by the marginal stability values at which the anisotropy should pin to. For the range of magnetizations we studied, the heating rate scales with shear frequency $s$ as approximately $s^{0.3}$ in the case of ions and $s^{0.2}$ in the case of electrons (fig. \ref{fig:EnergyGainScaling}).

On the other hand, simulation at lower $\beta$ (\S \ref{subsec:beta}) with higher mirror thresholds indicate that at too low a magnetization, the instability will not have time to develop at all, implying that there is a $\beta$ and shear amplitude dependent magnetization at which the heating is maximized. 

The alternating excitations  of mirror and firehose modes are essentially consistent with theoretical expectations in the kinetic regime (\cite{Pokhotelov2004,Hellinger2008}) and with previous studies of fully kinetic and hybrid simulations (\cite{Riquelme2015,Melville2016,Riquelme2018}). In the first half of the pump cycle ($0<t\cdot s<1$), we saw that mirror modes grow to relatively large amplitudes, have low frequencies and $kR_{L,i}\sim 1$ initially. After reaching nonlinear amplitudes, we also saw the appearance of short wavelength, parallel propagating modes in regions of low magnetic field with frequencies $\omega\sim 0.1 \omega_{c,e}^{\text{init}}$ and $kR_{L,i}\sim 1$, consistent with the nature of whistler lion roars (\cite{Baumjohann1999, Breuillard2018}). The effect of these modes on the global evolution and the nature of their interaction with the particles (e.g. resonant particle acceleration, \cite{Ley2019,Riquelme2017}) is deferred to future investigations (Ley et al. 2022 in prep.).

During the nonlinear mirror stage, the pressure anisotropy is regulated and maintained at a relatively constant amplitude above the marginally stable value, at least for the range of magnetizations studied here. In the second half of the pump cycle ($1<t\cdot s<2$), we saw the rapid excitation of low frequency, oblique firehose modes also with $kR_{L,i}\sim 1$, that quickly regulate the pressure anisotropy, maintaining it close to marginally stable values. Each subsequent cycle presents essentially the same evolution.


Even though the interaction of mirror and oblique firehose modes with the particles can become quite complex, especially in their nonlinear stages, we showed that a simplified model in which the effective pitch angle scattering frequency only depends on the growth rates of mirror and firehose modes when the approximate instability thresholds are met (similar to the bounded anisotropy model from \cite{Hellinger2008}) accurately reproduces the heating rate seen in our simulations. However, the model fails to capture the incomplete relaxation of the pressure anisotropy
to the linear instability threshold during the mirror dominated phase. This is evidence for an unidentified, possibly nonlinear damping process for the mirror modes, such that supercritical pressure anisotropy is needed to maintain them.

\cite{Lichko2017} have shown that a magnetic pumping configuration similar to\footnote{\cite{Lichko2017} used compression instead of shear, but other works have found little difference between these two types of pumping (\citep{Ley2019}).} the one studied in this work can create nonthermal tails in the distribution after several pump cycles, and their growth can be enhanced in the presence of particle trapping (\cite{Lichko2020}). Nonthermal tails appear on our simulations as well, and will be explored in forthcoming work (Ley et al. 2022, in preparation).

In the context of the ICM of galaxy clusters, we can expect to have a fully developed turbulent cascade with a large range of frequencies evolving simultaneously, possibly fed at high frequencies by cosmic rays. If we associate our local shear frequency $s$ with the frequency of the local turbulent motion, we can think of several channels of gyroviscous heating acting together, with different frequency-dependent efficiencies. Assuming that an anisotropy overshoot can be developed by any variation of the local magnetic field by a an ensemble of large-scale turbulent eddies, our results could constitute the building blocks to construct a plausible channel for turbulent dissipation in the ICM (\cite{Arzamasskiy2022}).

In this study, however, our results regarding the evolution of electrons are not complete, as our initial temperatures are too high to have a direct application to an nonrelativistic environment such as the ICM, especially for simulations with higher mass ratios. Electrons can also excite their own, electron-scale instabilities that can in turn interact with the electron population, contributing to the pitch-angle scattering, and they are not fully captured here.

We emphasize that the parameters considered here
represent tiny scales. For example, at a magnetization of 800,  if $B=0.1\mu G$, the time to complete a full pump cycle ($t\cdot s = 2$) is only $1.7\times 10^6$s. There is probably little turbulent power at such scales; otherwise the heating rate could easily become too large. In future work, we plan
to integrate our results with plausible turbulence models to
produce a full theory of gyroviscous heating in the ICM.

\begin{acknowledgments}
F.L.  acknowledges support from NSF Grant PHY-2010189 and the Fulbright Association. EZ acknowledges support from NSF Grant PHY-2010189 and the University of Wisconsin. MR thanks support from ANID Fondecyt Regular grant No. 119167. AT and LS were supported by NASA ATP 80NSSC20K0565. AT was supported by NASA FINESST 80NSSC21K1383. This work used the Extreme Science and Engineering Discovery Environment (XSEDE), which is supported by National Science Foundation grant number ACI-1548562. This work used the XSEDE supercomputer Stampede2 at the Texas Advanced Computer Center (TACC) through allocation TG-AST190019 (\cite{Towns2014}). This research was performed using the compute resources and assistance of the UW-Madison Center For High Throughput Computing (CHTC) in the Department of Computer Sciences. This research was partially supported by the supercomputing infrastructure of the NLHPC (ECM-02).
\end{acknowledgments}

\appendix
\section{Numerical Heating}
\label{sec:NumericalHeating}
The inherent discreteness of the macroparticles in PIC simulations introduces finite electric field fluctuations that can interact with the particles and affect the evolution of their physical properties. In our simulations, this effect introduces a numerical heating source in both ions and electrons that is comparable to their energy gain in a single pump cycle. In general, this numerical effect can be reduced by significantly increasing the number of particles per cell $N_{\text{ppc}}$, quickly making the simulations prohibitively expensive given our computational resources, especially for larger mass ratio $m_i/m_e$ and magnetizations $\omega_{c,i}^{\text{init}}/s$. In this section, we characterize this numerical heating in terms of its dependency on numerical and physical parameters and describe the method we applied to subtract it from the energy gain of each species.

\subsection{Weak Shear Simulations}
The characterization of the numerical heating was done by performing ``weak shear'' simulations, where the initial configuration is the same as the runs presented in previous sections (see fig. \ref{fig:PeriodicShear}$a$) but now reversing the shear at a much earlier time, $\tau_s=0.1 s^{-1}$ (see section \ref{sec:Simulation Setup}), so the variation in magnitude and direction of the magnetic field $\textbf{B}$ is so small that no significant pressure anisotropy is developed, no instability is excited and therefore the system does not exhibit any of the dynamics presented in the previous sections but the underlying numerical heating. This way, we can isolate the effect that the numerical heating produce upon the system.

\begin{figure}[hbtp]
    \centering
    \includegraphics[width=\linewidth]{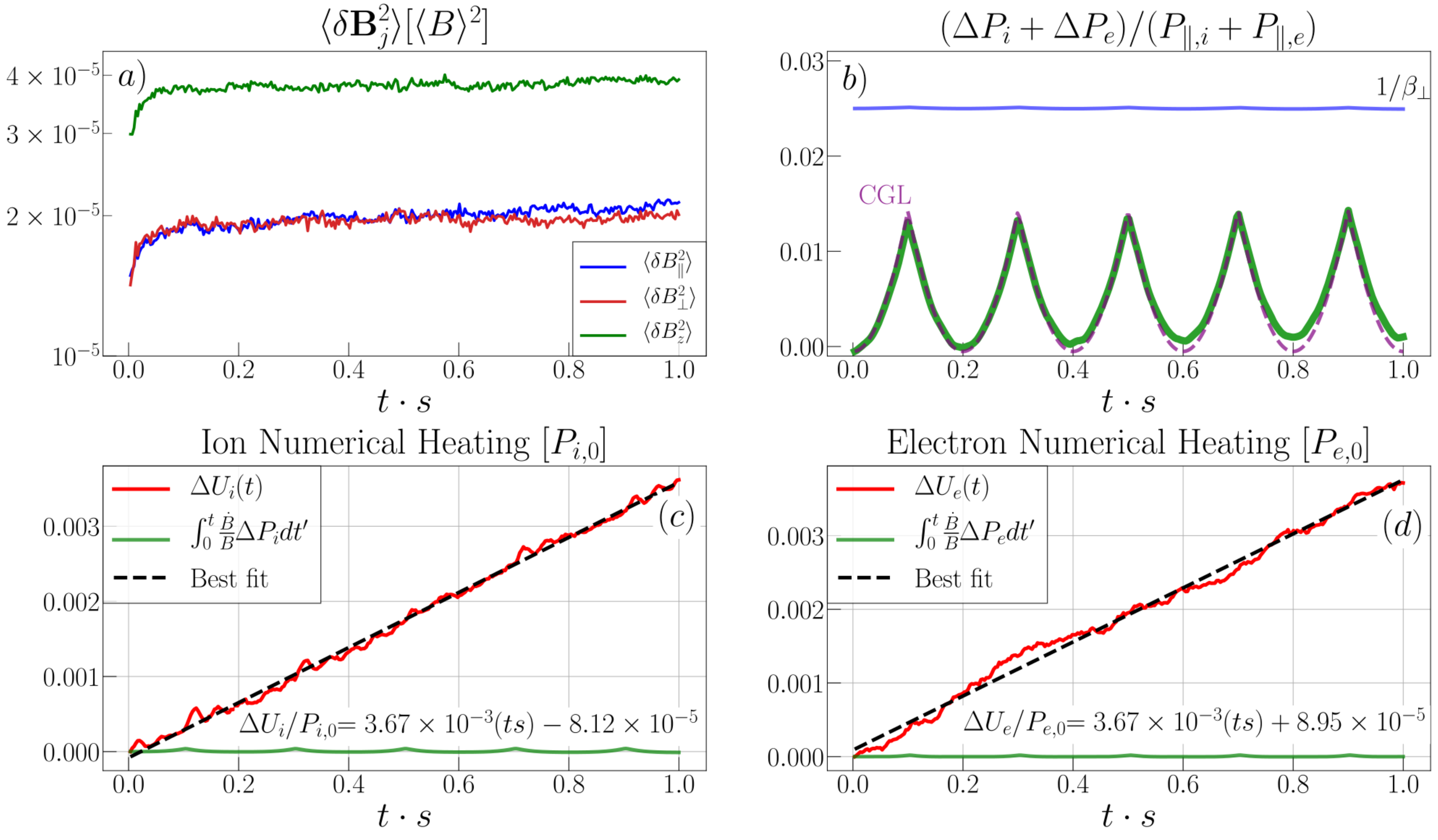}
    \caption{Weak shear simulation results for run Eb20m2w200 in table \ref{table:SimulationParameters}. Panel $a$: volume-averaged magnetic energy in $\delta \textbf{B}$ along different axes and as a function of time. Analogous to fig. \ref{fig:DeltaP}, here $\delta B_{\parallel}$ (blue line) is the component parallel to $\langle \textbf{B} \rangle$, $\delta B_{\perp,xy}$ (red line) and $\delta B_z$ (green line) are, respectively, the components perpendicular to $\langle \textbf{B} \rangle$ in the plane and perpendicular to the plane of the simulation. Panel $b$: The evolution of the total pressure anisotropy $\Delta P \equiv \Delta P_i+\Delta P_e$ (solid green line). The dashed purple line shows the corresponding double-adiabatic prediction (CGL, \cite{CGL1956}). The solid blue line shows the approximate linear threshold for the excitation of the mirror instability. Panel c: The solid red line shows the ion energy gain $\Delta U_i=U_i(t)-U_i(0)$ as a function of time; in solid green line it is shown the integrated ion gyroviscous heating rate and the dashed black line shows the linear best fit to $\Delta U_i$: $\Delta U_i/P_{i,0}=a_i(ts) + b_i$, where $a_i = 3.67\times 10^{-3} \pm 1.24\times 10^{-5}$ and $b_i = -8.12\times 10^{-5}\pm 7.2\times 10^{-6}$. Panel $d$: same as panel $c$ but for electrons, where now the best fit parameters to $\Delta U_e/P_{e,0}=a_e(ts)+b_e$ are $a_e=3.67\times 10^{-3} \pm 1.97\times 10^{-5}$ and $b_e=8.95\times 10^{-5} \pm 1.14\times 10^{-5}$.}
    \label{fig:WeakShear}
\end{figure}

Figure \ref{fig:WeakShear} shows the evolution of several physical quantities in one weak shear simulation, in which the reversal time is $\tau_s=0.1s^{-1}$(run Eb20m2w200 in table \ref{table:SimulationParameters}). We can see in fig. \ref{fig:WeakShear}$a$ that the energy in all components of the magnetic field fluctuations $\delta \textbf{B}$ stays constant at a low level throughout the simulation, so we confirm that no instability is developed. Similarly, in fig. \ref{fig:WeakShear}$b$ we see that the anisotropy $\Delta P/P_{\parallel}$ exhibits periodic variations due to the shear reversals, but always consistent with a double-adiabatic evolution (dashed purple line). It also never surpasses the threshold for mirror instability, consistent with the evolution of the magnetic field fluctuations. We can see the evolution of the energy gain $\Delta U_j(t) = U_j(t) - U_j(0)$ ($j=i,e$) for ions and electrons in solid red lines in figures \ref{fig:WeakShear}$c$ and \ref{fig:WeakShear}$d$, respectively, as well as the integrated gyroviscous heating rate for each species, shown in solid green lines. Consistent with the double-adiabatic evolution of the pressure anisotropy, the expected energization of ions and electrons by gyroviscosity evolves as periodic phases of heating and cooling but ultimately leading to zero net heating. Notwithstanding the above, the ion and electron energy gains clearly exhibit a steady growth that, in absence of any other physical heating source, reveals the action of numerical heating. In both figures \ref{fig:WeakShear}$c$ and \ref{fig:WeakShear}$d$, the dashed black line shows the best linear fit to $\Delta U_j$; $\Delta U_j = a_j(ts) + b_j$. These best fit relations are what we will be using to subtract the numerical heating from the particle energy gain in the rest of the simulations presented in this work.

\subsection{Dependency on Physical and Numerical Parameters}

We have shown that the numerical heating in our simulations grows linearly with time. Consequently, more heating will be accumulated for longer simulations. This is the case for higher mass ratios $m_i/m_e$ and higher magnetizations $\omega_{c,i}/s$. 

\begin{figure}[hbtp]
    \centering
    \includegraphics[width=\linewidth]{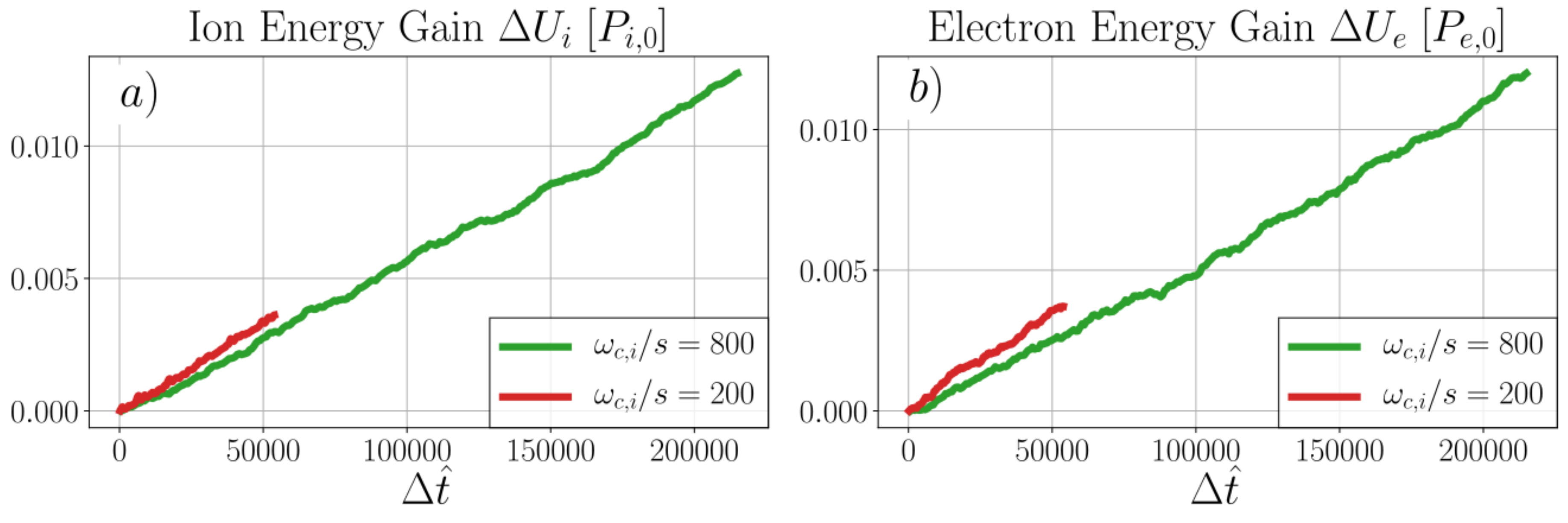}
    \caption{Panels $a$ and $b$ show, respectively, the ion and electron energy gain by the action of numerical heating as a function of time steps (not normalized) in the simulation for two runs with same $m_i/m_e=2$ and magnetization $\omega_{c,i}/s=200$ (Eb20m2w200, red line) and $\omega_{c,i}/s=800$ (Eb20m2w800, green line).}
    \label{fig:WeakShear_Magnetization}
\end{figure}

Figure \ref{fig:WeakShear_Magnetization} shows a comparison between the evolution of the ion and electron energy gains between two different weak shear runs (Eb20m2w200 and Eb20m2w800 in table \ref{table:SimulationParameters}) with $m_i/m_e=2$, $\tau_s=0.1s^{-1}$ and different magnetizations: $\omega_{c,i}^{\text{init}}/s=200$ (red line) and  $\omega_{c,i}^{\text{init}}/s=800$ (green line). In this case, the time shown in the horizontal axis is not normalized by their respective shear frequency $s$, but it shows the number of time steps in the simulation. In terms of normalized time, both runs Eb20m2w200 and Eb20m2w800 lasted until $t\cdot s=1$. With this choice, we can explicitly see that, even though the $\omega_{c,i}^{\text{init}}/s=800$ run is $\sim 4$ times longer, in both runs the numerical heating acts at a similar rate. 

\begin{figure}[hbtp]
    \centering
    \includegraphics[width=\linewidth]{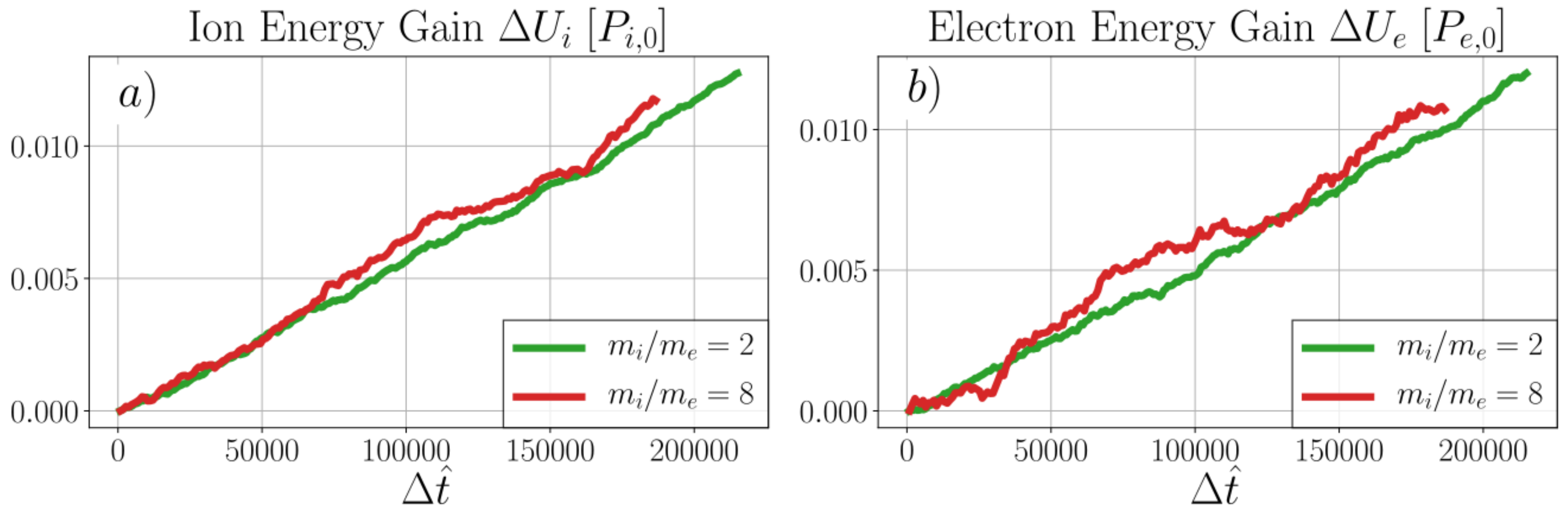}
    \caption{Panels $a$ and $b$ show, respectively, the ion and electron energy gain by the action of numerical heating as a function of time steps (not normalized) in the simulation for two runs with same $\omega_{c,i}/s=800$ and mass ratio $m_i/m_e=2$ (Eb20m2w800, red line) and $m_i/m_e=8$ (Eb20m8w800, green line).}
    \label{fig:WeakShear_MassRatio}
\end{figure}

The numerical heating behaves similarly also for different mass ratios, as shown in fig. \ref{fig:WeakShear_MassRatio}. In this case, we show the ion and electron energy gains for two weak shear runs (Eb20m2w800 and Eb20m8w800 in table \ref{table:SimulationParameters}) now with $\omega_{c,i}^{\text{init}}/s=800$, $\tau_s=0.1s^{-1}$ and different mass ratios: $m_i/m_e=2$ and $m_i/m_e=8$. We can see that both ions and electrons are numerically heated at a very similar rate as well, indicating that numerical heating is comparable among simulations with different mass ratios.


\subsection{Numerical Heating Subtraction}

Having characterized the nature of numerical heating in our simulations, we now show how it can be subtracted from the total energy gain in our periodic shear simulations. Given the linear evolution of the ion and electron energy gain by pure numerical heating, for every periodic shear run we show in this work, we performed a corresponding weak shear run with the same physical parameters and do a linear fit to the particle energy gain, similar to fig. \ref{fig:WeakShear}$c$ and \ref{fig:WeakShear}$d$. We then use the best-fit parameters thus obtained to subtract the numerical heating from the total energy gain in the periodic shear runs.

\begin{figure}[hbtp]
    \centering
    \includegraphics[width=\linewidth]{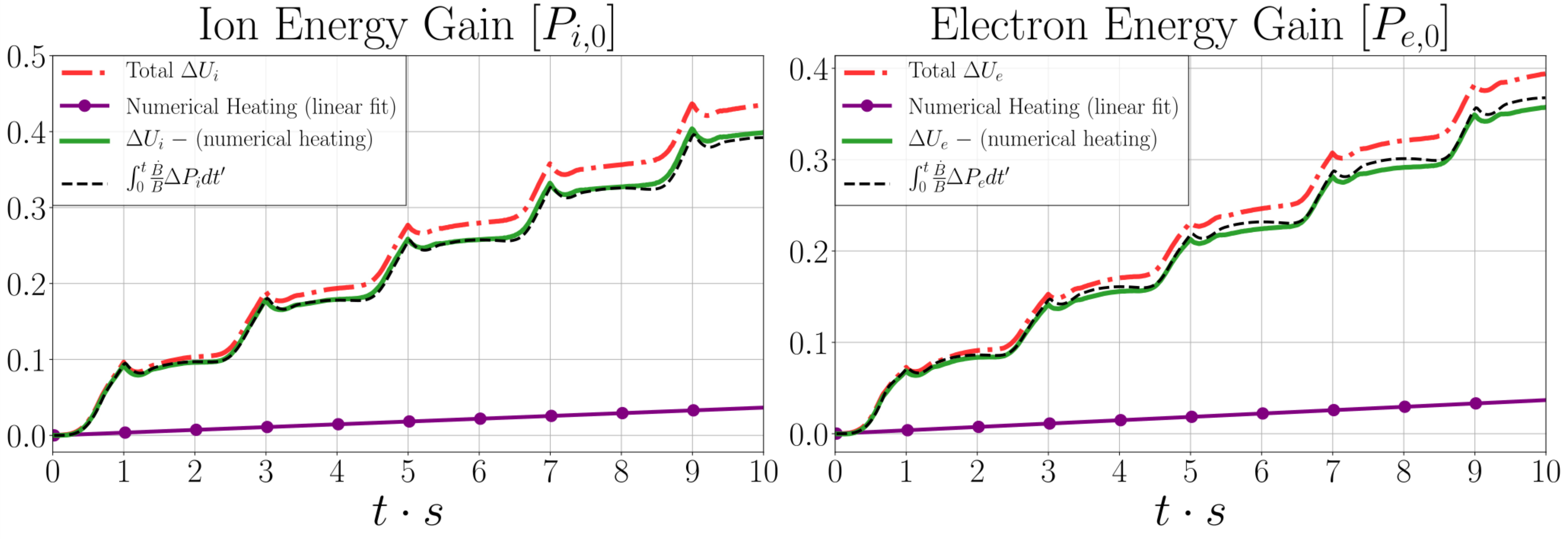}
    \caption{The ion and electron energy gain for run Zb20m2w200. Upper row: The dashed-dotted red line shows the evolution of the total ion energy gain including numerical heating. The solid purple line shows the linear best fit to the ion energy gain by pure numerical heating in a weak shear simulation (run Eb20m2w200, see fig. \ref{fig:WeakShear}$c$ and \ref{fig:WeakShear}$d$). The solid green line shows the subtraction of the numerical heating from the total energy gain using the linear best fit. Finally, the dashed black line shows the integrated ion gyroviscous heating rate. Bottom row: same as upper row but for electrons.}
    \label{fig:NumHeatSubtract}
\end{figure}

The results of this procedure are described in fig. \ref{fig:NumHeatSubtract} for ions (left panel) and electrons (right panel) for run Zb20m2w200. In both panels, the dashed red line shows the total energy gain of the particles, including the contribution from numerical heating. The solid purple line shows the linear best fit to the energy gain by pure numerical heating from the corresponding weak shear run Eb20m2w200 (dashed black lines from fig. \ref{fig:WeakShear}$c$ and \ref{fig:WeakShear}$d$). The numerical heating is then subtracted from the total particle energy gain using this linear fit, the result of which is shown in solid green line. We can see that it exhibits a better agreement with the integrated gyroviscous heating rate (dashed black lines).

\bibliographystyle{aasjournal}
\bibliography{main}

\end{document}